\documentclass[twocolumn]{aastex701}

\journalinfo{}          

\received{21 Aug 2025}
\accepted{03 Dec 2025}

\begin{document}

\title{Characterizing the Low-Mass Pre-Main-Sequence Population in the Low-Metallicity Star-Forming Region Dolidze 25 Using VLT-MUSE}

\author[]{Mizna Ashraf}
\affiliation{Department of Physics, Indian Institute of Science Education and Research Tirupati, Yerpedu, Tirupati - 517619, Andhra Pradesh, India}
\email{miznaashraf@gmail.com}  

\author[]{Jessy Jose} 
\affiliation{Department of Physics, Indian Institute of Science Education and Research Tirupati, Yerpedu, Tirupati - 517619, Andhra Pradesh, India}
\email{jessyvjose1@gmail.com}

\author[]{Gregory J. Herczeg}
\affiliation{Kavli Institute for Astronomy and Astrophysics, Peking University, China}
\email{gherczeg1@gmail.com}

\author[]{Min Fang}
\affiliation{Purple Mountain Observatory, Chinese Academy of Sciences, 10 Yuanhua Road, Nanjing 210023, China}
\email{mfang.cn@gmail.com}

\author[]{Varsha Ramachandran}
\affiliation{Astronomisches Rechen-Institut, Universität Heidelberg, Germany}
\email{vramachandran@uni-heidelberg.de}

\author[]{Carlo F. Manara}
\affiliation{ European Southern Observatory, Karl-Schwarzschild-Str. 2, D-85748 Garching bei München, Germany}
\email{cmanara@eso.org}

\author[]{Christian Schneider}
\affiliation{Hamburger Sternwarte, Gojenbergsweg 112, 21029 Hamburg, Germany}
\email{astro@pcschneider.eu}

\author[]{Megan Reiter}
\affiliation{Department of Physics and Astronomy, Rice University, 6100 Main Street – MS 108, Houston, TX 77005, USA}
\email{Megan.Reiter@rice.edu}

\author[]{Kiran Kumar Sunil}
\affiliation{University of Cumbria, Bowerham Road, Lancaster, Lancashire, LA1 3JD, United Kingdom}
\email{kiransnlkumar@gmail.com}

\correspondingauthor{Mizna Ashraf, Jessy Jose}
\email{miznaashraf@gmail.com, jessyvjose1@gmail.com}

\begin{abstract}

The metallicity of the star-forming environment is a fundamental parameter shaping the evolution of protoplanetary disks and the formation of planetary systems, yet its influence remains poorly constrained. We present a spectroscopic study of low-mass pre-main sequence (PMS) stars ($M < 1 \, M_\odot$) in the exceptionally metal-poor cluster Dolidze~25 ($Z \approx 0.2 \, Z_\odot$), using VLT/MUSE observations to probe accretion processes and disk evolution in a subsolar environment. We identify 132 cluster members using a combination of \textit{Gaia} astrometry and spectroscopic youth indicators, including lithium absorption and Balmer emission. The stellar parameters are derived using low-metallicity BT-Settl models yielding effective temperatures, extinctions, luminosities enabling robust estimates of stellar masses and ages. Mass accretion rates ($\dot{M}_\mathrm{acc}$) derived from H$\alpha$ emission span $10^{-10}$--$10^{-8} \, M_\odot\,\mathrm{yr}^{-1}$ with a median value of \(8 \times 10^{-10}\,M_\odot\,\mathrm{yr}^{-1}\). These rates are comparable to those in solar-metallicity regions of similar age, such as Lupus and Orion, indicating minimal metallicity dependence in accretion processes. Our analysis shows that using solar-metallicity templates to fit low-metallicity stars leads to systematic overestimations of \(T_\mathrm{eff}\) (by approximately \(300\,\mathrm{K}\)) and \(A_V\) (by around \(0.5\,\mathrm{mag}\)), underscoring the importance of employing metallicity-matched models for reliable characterization in low-\(Z\) environments. We present flux-calibrated, extinction-corrected spectra of these metal-poor PMS stars as a valuable resource for future investigations of disk evolution in subsolar regimes.

\end{abstract}


\section{Introduction} 

Low-mass stars form through the gravitational collapse of dense cores within molecular clouds, giving rise to a central protostar surrounded by a protoplanetary disk. Mass transfer from the disk onto the pre-main sequence star occurs predominantly through magnetospheric accretion, a process whereby material is channeled along the stellar magnetic field lines \citep{2007Bouvier, 2012Alencar}. This accretion process plays a crucial role in regulating stellar mass growth and drives the evolution and eventual dispersal of the circumstellar disk \citep[see review by][]{Hartmann2016}.

The rate of mass accretion and the timescale of disk dispersal are influenced by a combination of physical and environmental factors, including the mass of the host star \citep{Manara2022}, the metallicity of the star-forming region \citep{2014Tanaka}, stellar density \citep{2016zwart}, and external photoevaporation \citep{2018Nakatani,2024Damian}. The initial metallicity of the disk, inherited from the parent molecular cloud, is predicted to affect disk dispersal timescales by altering the dust-to-gas ratio, disk opacity, and the efficiency of mass accretion and photoevaporation\citep{2019Wolfer,2023Oberg}. As circumstellar disks provide the raw material for planet formation, their evolution and eventual dissipation have a critical impact on the overall potential for planet formation \citep[see reviews by][]{2011Williams, Andrews2020, Miotello2022}.

Theoretical models provide divergent predictions on accretion dynamics and disk dispersal as a function of metallicity. In metal-poor environments, higher gas temperatures and hence higher ionization fractions may lead to enhanced accretion rates, while simultaneously reducing disk lifetimes \citep{2009Hosokawa,2010Ercolano,2013Dopcke}. The reduced dust content in metal-poor disks allows far-ultraviolet (FUV) radiation to penetrate deeper, thereby enhancing photoevaporation rates and accelerating disk dispersal \citep{2009Gorti,2016Bai,2021Ercolano,2023Gehrig}. Additionally, lower metallicities limit radiative cooling via metal lines, suppressing cloud fragmentation, potentially resulting in higher average stellar masses \citep{2001Kroupa}. Alternatively, higher gas temperatures may also lead to higher disk masses, which could lengthen, rather than shorten, disk lifetimes.

Observational studies of low-metallicity protoplanetary disks remain sparse and often inconclusive. This limitation arises primarily because most low-metallicity star-forming regions are located at large distances, typically in the outer regions of the Milky Way or in nearby dwarf galaxies such as the Large and Small Magellanic Clouds (LMC and SMC). These distances hinder the detailed characterization of individual YSOs and limits the use of reliable accretion diagnostics, such as UV continuum and emission line measurements. 

Photometric excess based estimates of YSO accretion rates in the SMC and LMC have reported systematically higher values relative to solar-metallicity environments \citep[e.g.,][]{2011DeMarchi,2012Spezzi,2013Demarchi,2017DeMarchi,2019Biazzo,2023Tsilia,2023Vlasblom}. However, contrasting findings have been reported in Galactic low-metallicity clusters. \citet{Kalari2015} found accretion rates in Sh~2-284 ($Z \sim 0.2\,Z_\odot$) and \citet{2023Itrich} for the Canis Major region ($Z \sim 0.55\text{--}0.73\,Z_\odot$) to be comparable to those observed in solar-metallicity environments using spectroscopic tracers. The impact of metallicity on disk fractions also remains debated. Studies using NIR excess report lower disk fractions in low-metallicity clusters of similar age, implying faster disk dispersal \citep{2010Yasui, Guarcello2021, 2021Yasui, 2024Patra}, while \citet{2024DeMarchi} argue from H-alpha measurements that accretion may persist for longer in metal-poor environments.

In this context, the young cluster Dolidze 25 offers a particularly valuable case study. It is the closest known low-metallicity star-forming region, located at a Galactocentric radius of approximately $\sim12$~kpc and with a metallicity of $Z \sim 0.2 \, Z_\odot$ \citep{Lennon1990, Guarcello2021, Kalari2015, 2024Patra}. The relative proximity of the cluster allows for detailed spectroscopic characterization of YSOs under metal-poor conditions.

In this study, we use optical spectroscopy from VLT/MUSE to conduct a detailed characterization of low-mass pre-main sequence (PMS)  stars in the central region of the Dolidze 25. Previous investigations by \citet{Kalari2015} and \citet{2011Cusano} primarily focused on the relatively massive stellar population (median stellar mass \(\sim 1.6\,M_\odot\)) of the cluster. Our analysis targets the previously unexplored lower-mass population (median mass \(\sim 0.3\,M_\odot\)). We evaluate their accretion properties, masses, and other fundamental parameters, comparing our results with those from solar-metallicity star-forming regions to assess the impact of low metallicity on accretion and disk evolution.

The paper is organized as follows: Section~\ref{sec:observations} describes the VLT/MUSE observations, data reduction, and the archival data used in this work. Section~\ref{sec:dolidze25} outlines the properties of Dolidze 25, including its distance and metallicity estimates. Section~\ref{sec:analysis} details the spectral analysis and pre-main sequence star identification. Section~\ref{sec:results} presents the results, including derived stellar parameters and accretion rates. Section~\ref{sec:discussion} discusses the role of environment in regulating mass accretion rates, as well as the potential limitations of using solar-metallicity templates to characterize low-metallicity stellar spectra. Section~\ref{sec:conclusion} summarizes the main conclusions of this study.

\begin{figure*}
\centering
\includegraphics[width=0.8\textwidth]{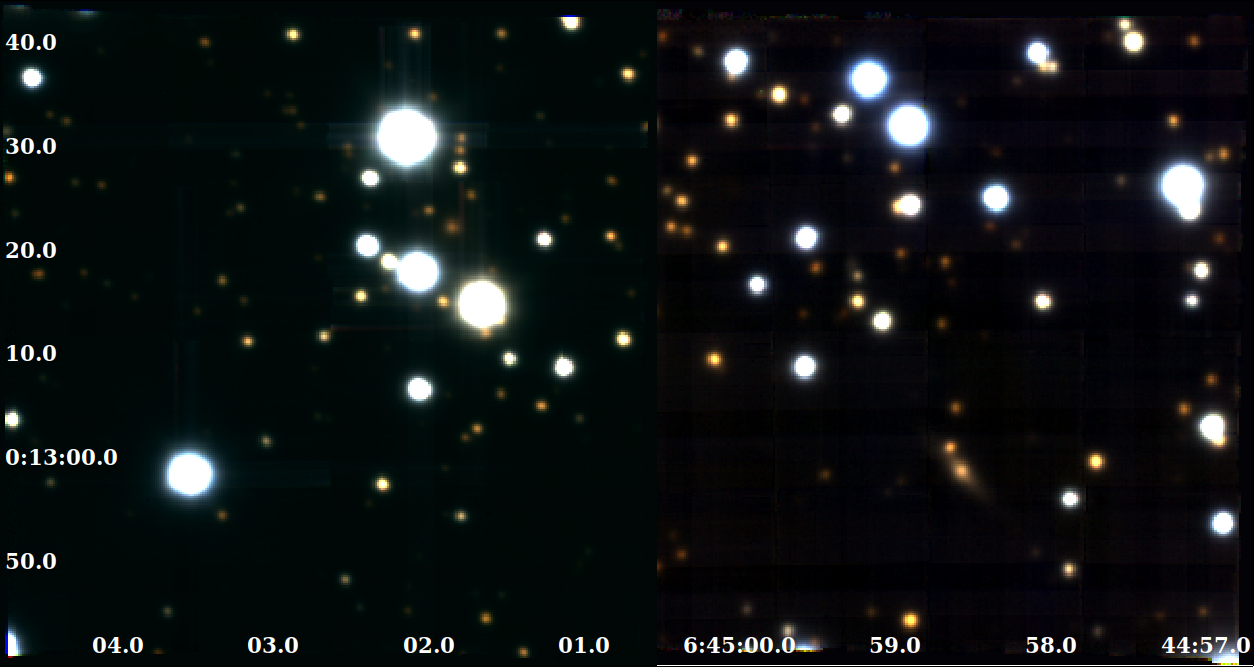}
\caption{RGB image of Dolidze 25 created from the \(r\)-, \(i\)-, and \(z\)-band images generated by integrating the MUSE datacube over the corresponding Pan-STARRS filter wavelength ranges. Two pointings were obtained toward the cluster core, centered at (06\textsuperscript{h}45\textsuperscript{m}02\fs68, +00\degr13\arcmin11\farcs4) for Field~C (left) and (06\textsuperscript{h}44\textsuperscript{m}58\fs82, +00\degr13\arcmin11\farcs4) for Field~D (right), covering the central \(2'\times1'\) region of the cluster.}
\label{fig:muse}
\end{figure*}

\begin{figure*}
\centering
\includegraphics[width=0.99\textwidth]{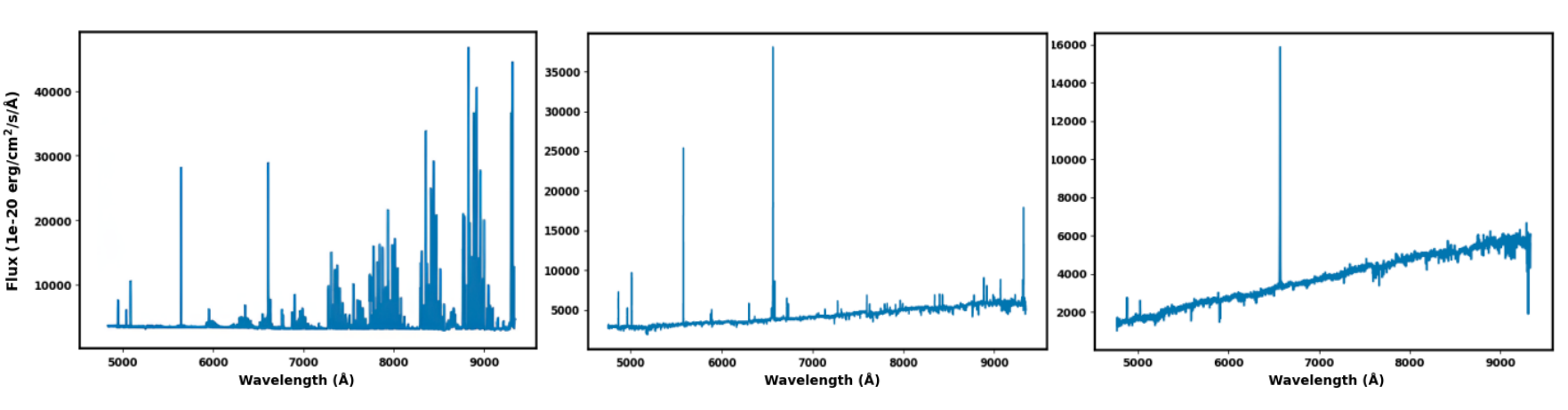}
\caption{Left: Sample spectrum without sky subtraction, processed with ESORex, showing telluric and nebular contamination from a background region. Middle: Spectrum of a point source from \textsc{MUSEpack}-reduced cube with telluric absorption removed while retaining the nebular emission background. Right: Final spectrum after subtracting both telluric and nebular emission, revealing the intrinsic spectrum of the point source.}
\label{fig:contaminants}
\end{figure*}

\section{Observations and Archival Data} \label{sec:observations}

The primary dataset was obtained from VLT-MUSE, accessed through the ESO Science Archive. To complement our analysis, we incorporated archival photometry from GAIA DR3 \citep{2023Gaia} and Pan-STARRS DR1 \citep{2016Chambers}. Additionally, we retrieved UVES-ECHELLE spectra (Program ID: 384.B-0066; PI: Nieva, Maria-Fernanda) of the central O-type star and FLAMES-GIRAFFE spectra (Program ID: 082.D-0839; PI: Neiner, Coralie) of both the central O-type star and a nearby B-type star from the ESO Science Archive.

\subsection{MUSE Observations and Data Reduction} \label{subsec:muse}

Observations of the central \(2'\times1'\) area of the Dolidze 25 star-forming region were carried out using VLT/MUSE under ESO program 098.C-0435 (PI: G. Herczeg). The observations were obtained in the nominal wavelength range of 4750--9350~\AA\ using the wide-field mode (WFM-NOAO) of the instrument, which covers \(1'\times1'\) per pointing. The spectral resolution of MUSE ranges from \(R \sim 1750\) at 4750~\AA\ to \(R \sim 3750\) at 9350~\AA, with a resolving power of \(\sim 2500\) at H$\alpha$ (6563~\AA). The 5$\sigma$ point-source sensitivity reaches approximately $\sim 24.2$~mags in $V_{\mathrm{AB}}$ band. Two pointings were obtained towards the core of the cluster centered at (06:45:02.68, +00:13:11.4); Field C and (06:44:58.82, +00:13:11.4); Field D (Figure~\ref{fig:muse}). The atmospheric seeing during the observations was approximately 0.7$^{\prime\prime}$, which is well sampled by the MUSE spatial pixel scale of 0.2$^{\prime\prime}$ per pixel.

We used \textsc{MUSEpack}, a Python-based wrapper to the EsoRex pipeline \citep{Peter2019} to reduce the raw data. \textsc{MUSEpack} improves sky subtraction in crowded fields and regions with strong, non-uniform nebular emission by modifying the sky model used in the pipeline. Specifically, the sky lines template is modified to retain only telluric emission lines (OH and O$_2$), excluding astrophysical lines such as the Balmer series, [N\,\textsc{ii}], and [S\,\textsc{ii}], to ensure that only atmospheric contributions are subtracted. Additionally, the sky continuum subtraction is set to zero under the assumption that the continuum of the H\,\textsc{ii} region dominates over the telluric continuum. This approach ensures that only atmospheric emission is subtracted, preserving the astrophysical signal in the final datacube. The modified data reduction with MUSEpack successfully removed telluric contamination while retaining the intrinsic morphology and flux of nebular emission lines (see Figure~\ref{fig:contaminants}), providing a datacube suitable for further analysis.

\begin{figure*}
\centering
\includegraphics[width=0.9\textwidth]{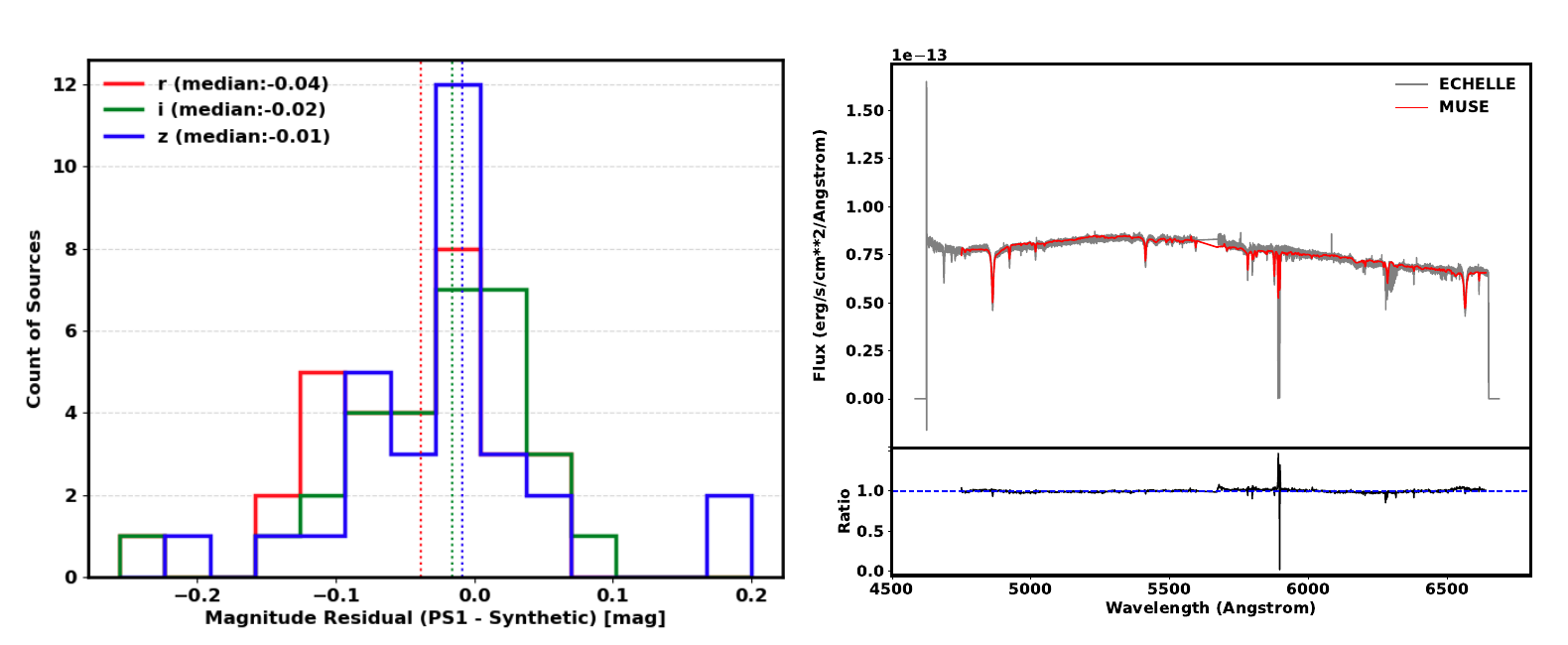} 
\caption{Left: Comparison of synthetic photometry from MUSE with Pan-STARRS , where the residual is defined as \textit{Pan-STARRS $-$ Synthetic Photometry}. The median residual values are calculated from a sample of 30 sources, with Pan-STARRS $i$ magnitudes brighter than 19 and $r$ magnitudes brighter than 20. Right: Comparison of MUSE spectra for the brightest source in the field of view with high-resolution VLT-UVES spectra, showing consistent flux calibration between the two.}
\label{uves}
\end{figure*}

\subsubsection{Spectral Extraction and Nebular Background Correction} \label{sec:methodology}

The full width at half maximum (FWHM) of the point spread function, corresponding to the atmospheric seeing of $\sim$0.7$^{\prime\prime}$, spans approximately 4 spatial pixels in the MUSE data. Circular apertures are typically defined as multiples of the FWHM, with a radius of 1~FWHM enclosing approximately 76\% of the total flux. Apertures with radii of 3–4~FWHM are generally required to recover nearly the entire flux of a point source. Given the high stellar density in Dolidze 25, such large apertures are not practical. Therefore, we extracted spectra using a circular aperture with a radius of 4 pixels (corresponding to $\sim$1~FWHM) and applied aperture corrections to account for the missing flux outside this radius.

The extraction of stellar spectra from the MUSE datacube was performed using the \texttt{ivar} (inverse variance) method provided by the \texttt{PyMUSE} package \citep{2020Pessa}. This technique weights each spaxel by the inverse of its variance, thereby minimizing noise contributions from low-S/N pixels and optimizing the signal-to-noise ratio (SNR), especially for faint sources. Spectrum of point sources were extracted within a circular aperture of 4-pixel radius, and normalized to match the integrated flux obtained using the sum method to ensure consistency between inverse variance and sum methods.

For nebular background subtraction, we adopted a modified version of the method described by \citet{2024Rogers}. For each source, a local background spectrum was extracted from an annular region with inner and outer radii of 20 and 30 pixels, respectively, ensuring that the background is located at least 4~FWHM away to minimize flux contribution from the source itself. To mitigate contamination from nearby stars, all sources falling within the annulus were masked from the datacube with a radius of 3 FWHM (12 pixels). The median background within this annulus was calculated and scaled to match the aperture area of the source. Due to the strong spatial variability of nebular emission across the H\,\textsc{ii} region, direct subtraction of the local background often led to over-subtraction, resulting in unrealistic negative fluxes. To address this, we implemented a line-based scaling approach in which the background spectrum was iteratively scaled to minimize the combined equivalent widths (EWs) of the [N\,\textsc{ii}]~$\lambda6584$ and [S\,\textsc{ii}]~$\lambda6731$ forbidden lines in the resulting, background-subtracted spectrum.

This method relies on the assumption that [N II] and [S II] emission in low-mass PMS stars is dominated by the surrounding nebula, with negligible intrinsic stellar contribution. Observations in nearby low-mass star-forming regions (e.g., Lupus, Orion, and Upper Scorpius) show that emission from [N\,\textsc{II}] and [S\,\textsc{II}] is generally weak or absent, and when detected, it is often spatially extended or linked to disk winds \citep{2014Natta, 2023Gangi, 2023Fang}.

The Balmer emission lines observed in background regions are found to correlate with the strength of [N\,\textsc{ii}] and [S\,\textsc{ii}] (see Fig.~\ref{fig:nii_sii_halpha} in Appendix); thus, scaling to remove these forbidden lines effectively eliminates the associated nebular Balmer contamination. The median scaling factor obtained is 0.95, indicating a slight underestimation of the stellar flux or an overestimation of the background. The resulting line-scaled background subtraction yielded cleaner H$\alpha$ profiles and significantly reduced the occurrence of unphysical negative fluxes.

A potential caveat of this method is the presence of circumstellar [N\,\textsc{ii}] or [S\,\textsc{ii}] emission originating from accretion disks, proplyd-like structures, jets, or outflows, which could lead to oversubtraction if these lines are minimized to zero \citep{2024Aru,2025Mauco}. However, since most of the background spectra are scaled down (median scaling factor $\sim$0.95), the occurrence of this effect is rare. In cases of oversubtraction, the measured H$\alpha$ flux could be reduced, potentially leading to an underestimation of the accretion rate by up to 10\%. A detailed explanation of the scaling procedure, including estimates of spatial variability and correlations between line strengths, is presented in Appendix~\ref{app:nii_subtraction}.

\subsubsection{Aperture correction}

To account for the flux loss due to using a small aperture (1~FWHM, or 4 pixels in radius), we derived aperture correction factors based on empirical measurements from isolated stars. A sample of isolated sources—17 in Field~D and 10 in Field~C—was selected to estimate the scaling factor. For each source, spectra were extracted using circular apertures with radii of 1~FWHM (4 pixels) and 4~FWHM (16 pixels) using \texttt{PyMUSE}, following the procedure outlined in Section~\ref{sec:methodology}. The correction factor was calculated as the wavelength-dependent ratio of the flux in the 4~FWHM spectrum to that in the corresponding 1~FWHM spectrum. 

These ratios vary slightly with wavelength due to the PSF profile and instrumental response; however, the overall trend was consistent across sources within each field. Therefore, the median ratio across all isolated stars in each field was adopted as the aperture correction function and applied multiplicatively to all 1~FWHM spectra of the respective field. This ensures that the extracted fluxes are accurately calibrated and comparable to those obtained using larger apertures, without introducing biases due to crowding or variable seeing conditions.

\subsubsection{Calibration Accuracy}

The accuracy of flux calibration was evaluated by performing synthetic photometry in the  $r$, $i$, and $z$ bands for sources within the MUSE field of view that have Pan-STARRS photometry available in the archive. To minimize errors and exclude spurious sources, only those with $i$-band brighter than 19 and $r$-band brighter than 20 were included, resulting in a sample of 30 sources. The synthetic photometry were computed by integrating the source spectra over the Pan-STARRS filter curves using the pyphot library \citep{pyphot}. For each band, the difference between the pan-STARRS and synthetic photometry were calculated, and median offset were determined. The median offsets for all bands were found to be minimal, with values less than 0.05 (r: 0.04, i: 0.02, z: 0.01), indicating strong agreement with the Pan-STARRS magnitudes. The distribution of residuals across the bands is shown in Figure \ref{uves}. These results demonstrate that the synthetic photometry is in good agreement with the Pan-STARRS magnitudes, with minimal offsets, confirming the reliability of the flux calibration within the specified magnitude limits. Although source variability could introduce discrepancies between the MUSE and Pan-STARRS fluxes, such effects are expected to average out over the sample and do not significantly impact the overall result.

 To further confirm the accuracy of flux calibration, we compared the aperture-corrected MUSE spectrum with the high-resolution UVES spectrum of the central O-type star (S15: \(\alpha, \delta = 101.2591, 0.2246\)). The flux-calibrated UVES Echelle spectrum for the central O-type star was obtained from the ESO Science Archive. It has a spectral resolution of \(R = 34{,}540\), covers the range 4583–6687~\AA, and has a signal-to-noise ratio (SNR) of 169. Flux calibration was performed by the ESO Quality Control Group using master response curves derived from standard star observations. To assess the accuracy of the aperture correction, the MUSE and UVES spectra were compared in their overlapping wavelength region (see Fig.~\ref{uves}); the median absolute deviation is close to zero, indicating excellent agreement.

\begin{figure}[htbp]
\centering
\includegraphics[scale=0.35]{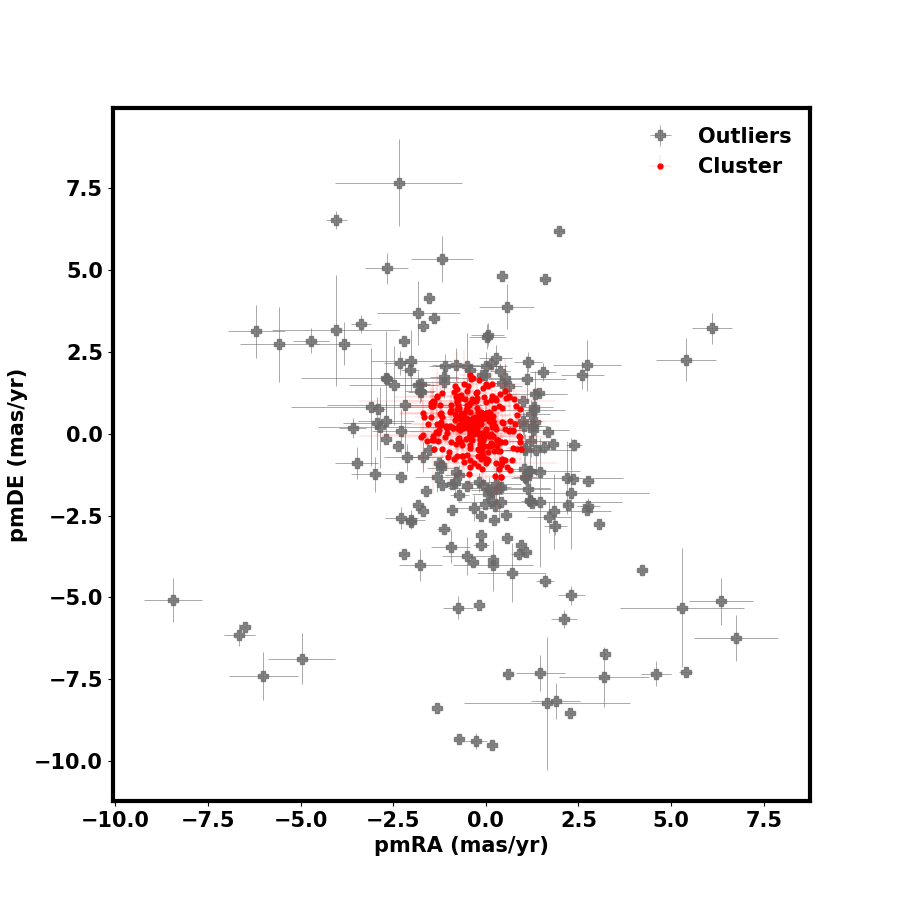}
\caption{Proper motion in Right Ascension (pmRA) versus Declination (pmDec) for sources within a 3$\arcmin$ radius of Dolidze~25. Red points represent cluster members identified by DBSCAN; gray points indicate outliers.}
\label{fig:cluster}
\end{figure}

\begin{figure*}[htbp]
\centering
\includegraphics[scale=0.44,trim={0cm 0cm 0cm 6.5cm},clip]{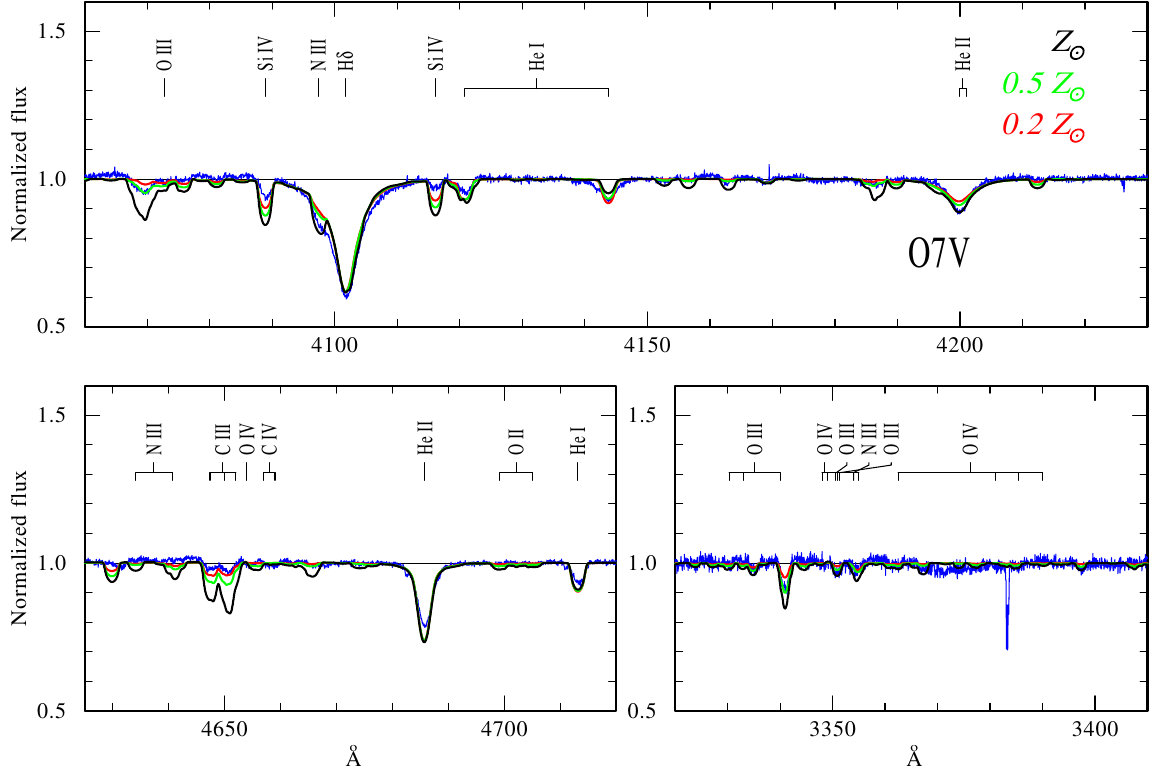}
\includegraphics[scale=0.44,trim={0cm 0cm 0cm 6.5cm},clip]{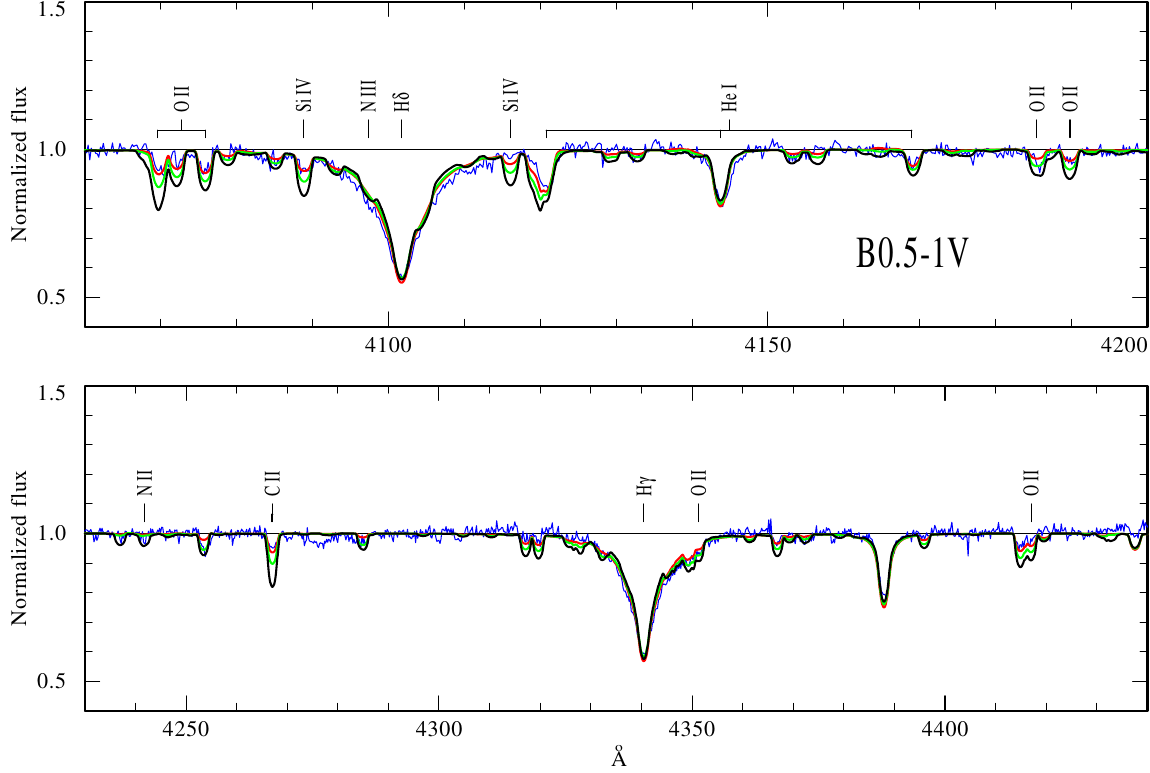}
\caption{Comparison of observed optical spectra (blue) of Cl* Dolidze 25 MV 15 (O7\,V) and Cl* Dolidze 25 MV 18 (B0.5–1\,V) with synthetic spectra from PoWR models at three different metallicities: solar ($Z/Z_\odot = 1.0$, black), LMC ($Z/Z_\odot = 0.5$, green), and SMC ($Z/Z_\odot = 0.2$, red). Key diagnostic lines, including He, CNO, and Si, are marked. The observed weakness of metal lines relative to the solar model indicates a subsolar metallicity.}
\label{fig:OBmetallicity}
\end{figure*}

\section{Dolidze 25: Cluster Properties and Parameters}\label{sec:dolidze25}

Dolidze~25 \( (l = 211.9^\circ, b = -1.3^\circ) \) is one of the best known rare cases of Galactic low-metallicity environments, with a commonly adopted distance of \(\sim 4.5~\mathrm{kpc}\) based on previous studies \citep{Delgado2010,2011Cusano}. It is a young \citep[1 -2 Myr;][]{2024Patra} open cluster associated with the H II region Sharpless 2-284 located towards the Galactic anticentre. Spectroscopic analyses of early-type stars in the Sh2-284 region reveal significantly subsolar metallicities, with silicon and oxygen abundances approximately $0.3\,\mathrm{dex}$ and $0.5\,\mathrm{dex}$ below solar, respectively \citep{Negueruela2015}. These findings, in agreement with earlier studies \citep{Lennon1990, Fitzsimmons1992}, confirm Dolidze 25 as one of the most metal-poor star-forming regions identified in the Milky Way \citep{Kalari2015}. The cluster hosts a population of massive OB stars that contribute to the ionization of Sh2-284 \citep{Negueruela2015}. With its unique interplay of low metallicity, intense radiation fields, a young stellar population, and proximity, Dolidze 25 offers an exceptional laboratory for investigating the formation and evolution of stars and planetary systems in metal-poor environments.

\subsection{Distance Estimation} \label{sec:distance}

We estimated the distance to Dolidze~25 to be \(4.2 \pm 0.5\)~kpc, based on the parallax of cluster members identified from the \textit{Gaia}~DR3 catalog \citep{2023Gaia} for a region much larger than the MUSE field of view. Within the MUSE FOV, only a few stars have reliable \textit{Gaia} parallaxes; therefore, to obtain a robust distance estimate, we selected a larger area centered on the MUSE field and identified cluster members. \textit{Gaia}~DR3 was queried for sources within a 3\arcmin\ radius of the central region of Dolidze~25, yielding 504 objects with proper motion (\(\mu_{\alpha}, \mu_{\delta}\)) and parallax measurements. To ensure high-precision astrometry, only sources with a renormalized unit weight error (RUWE) below 1.4 were retained.

We used the DBSCAN (Density-Based Spatial Clustering of Applications with Noise) algorithm with parameters \texttt{eps=0.5} and \texttt{min\_samples=25} on the proper motion data (\(\mu_{\alpha}\), \(\mu_{\delta}\)). The \texttt{eps} parameter sets the maximum separation of 0.5 milliarcsecond per year for two points to be considered neighbors. The \texttt{min\_samples} parameter specifies the minimum number of points required to form a dense region. This allowed us to identify sources that formed a cluster and exclude outliers, as shown in Figure \ref{fig:cluster}. Further, sources with parallaxes deviating by more than \(5\sigma\) from the weighted mean parallax of the cluster were excluded as outliers, with weights assigned as \(1/\sigma^2\).

After applying these criteria, 258 sources were confirmed as probable cluster members within the larger 3\arcmin area , and their weighted mean parallax yielded the distance estimate, with the uncertainty given by the standard deviation of the parallax distribution.

\subsection{Metallicity Estimation} \label{sec:metallicity}

We performed a detailed spectroscopic analysis to estimate the metallicity of two massive main-sequence stars in the Dolidze 25 cluster: Cl* Dolidze 25 MV 15 (O7 V) and Cl* Dolidze 25 MV 18 (B0.5–1 V). For the O7 star, we utilized high-resolution UVES spectra retrieved from the Very Large Telescope (VLT) archive, while for the B-type star we used available medium-resolution FLAMES spectra.

To determine the metallicity ($Z$), we compared the observed spectra with synthetic spectra generated using the non-LTE Potsdam Wolf-Rayet (PoWR) model\footnote{http://www.astro.physik.uni-potsdam.de/PoWR/} atmosphere code \citep{Grafener2002, Hamann2003, Sander2015}. As a starting point, we selected models from the Milky Way grid with solar metallicity ($Z/Z_\odot = 1.0$), as well as lower metallicity models from the LMC ($Z/Z_\odot \approx 0.5$) and SMC ($Z/Z_\odot \approx 0.2$) grids \citep{Hainich2019}. The PoWR models include detailed model atoms for H, He, C, N, O, Mg, Si, P, S, and the iron group elements (using a superlevel approach). While their overall abundances are scaled with the corresponding metallicity, the Fe abundance cannot be directly constrained from optical spectra due to the lack of strong diagnostic lines in hot OB stars. Therefore, our metallicity estimates primarily depend on observable CNO and Si line features.

The basic stellar parameters, such as effective temperature ($T_\mathrm{eff}$) and surface gravity ($\log g$), were estimated by fitting the synthetic spectra to the observed optical spectra, focusing on the helium line ratios and the wings of the Balmer lines. The projected rotational velocity ($v \sin i$) was determined from the broadening of metal lines. For more details on our spectral analysis methodology, see the approach described in \citet{Ramachandran2018b, Ramachandran2019}.

After constraining the fundamental stellar parameters, we compared models with the three different metallicities as shown in Fig.\,\ref{fig:OBmetallicity}. The strengths of the C, N, O and Si absorption lines in the observed spectra are significantly weaker than those predicted by the solar metallicity models. Almost all of the metal absorption features are best reproduced by the SMC-like metallicity models, with some lines matching the LMC models. This suggests that the average metallicity of the Dolidze 25 region is comparable to that of the SMC ($Z/Z_\odot \approx 0.2$).

\section{MUSE Spectral Analysis} \label{sec:analysis}

This section outlines the methodology and presents the results of the analysis performed on the MUSE spectra, aimed at identifying and characterizing the pre-main-sequence stellar population in Dolidze~25. A total of 287 spectra were extracted from the datacube by the method presented in Section \ref{subsec:muse}, of which 132 with signal-to-noise ratios (SNR) $> 5$ (median SNR between \(\lambda6540\text{-}6550\,\text{\AA}\)) were selected for further analysis, providing sufficient quality for identifying spectral features and carrying out quantitative fitting.

Spectral types were determined for all good-SNR sources (SNR $> 5$), regardless of their status as cluster members or field stars. We identified PMS candidate members based on spectroscopic youth indicators, such as Li\,\textsc{i} \(\lambda6708\) absorption or signatures of active accretion, and analyzed their properties. The spectral fitting methods implemented, along with the specific criteria and techniques used for the identification and classification of PMS stars, are described in the following sections.

\begin{figure*}[htbp]
\centering
\includegraphics[width=0.9\textwidth]{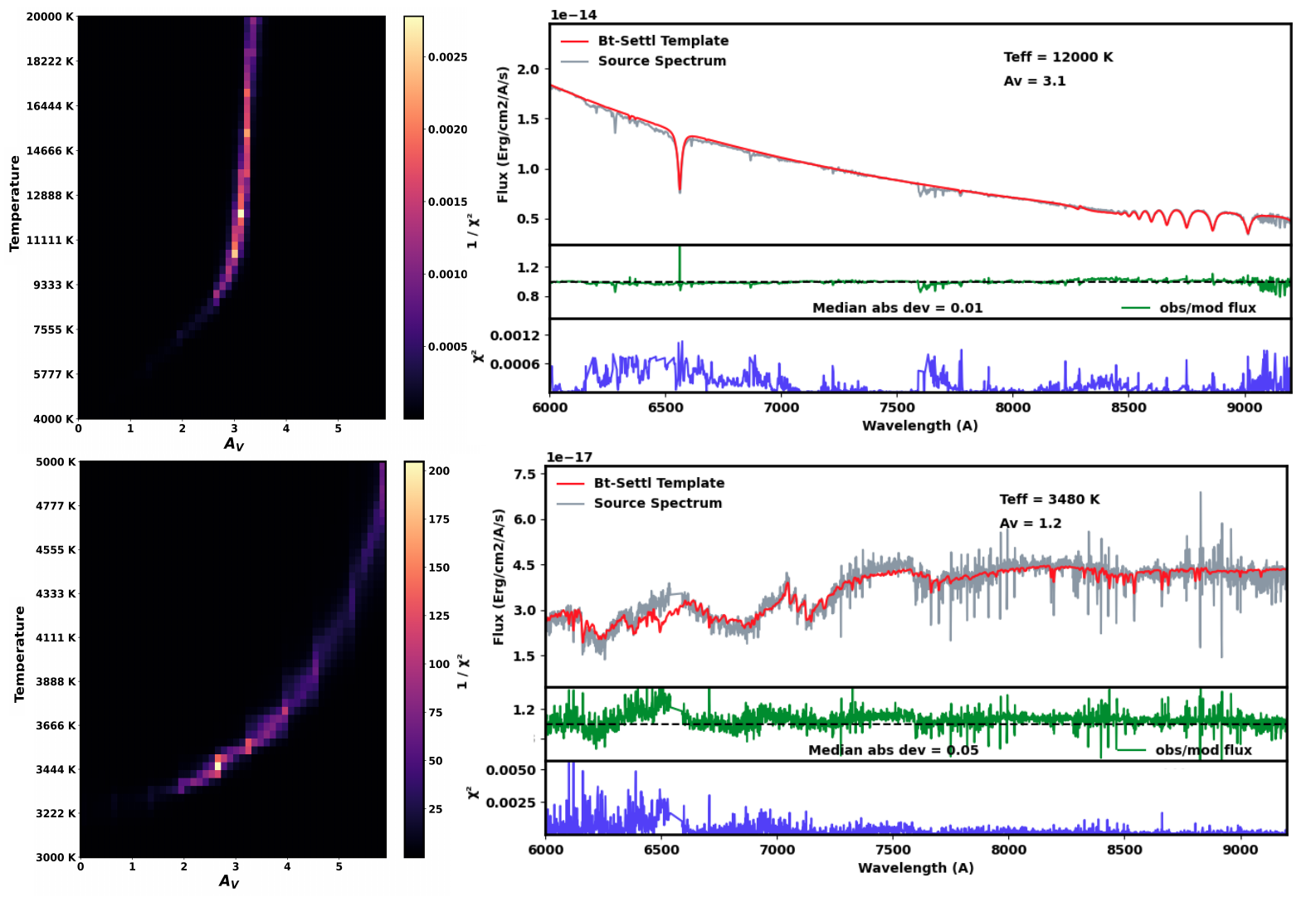} 
\caption{\textbf{Left:} Hess diagram of the $1/(\chi^2)$ distribution for a B-type (top) and M-type (bottom) source. The brightest pixel corresponds to the best-fitting template, with \(A_V\) as a variable. \textbf{Right:} The best-fit template for the same source obtained from the \(\chi^2\) minimization between the source and template within the wavelength range of 6500–9500 \AA, assuming a constant metallicity of \(-0.7\), \(R_V = 3.1\)and variable \(A_V\), \(\log(g)\) and ($T\textsubscript{eff}$) (top panel). The middle panel shows the ratio between the dereddened observed flux and the scaled template flux, along with its median absolute deviation. The bottom panel displays the \(\chi^2\) calculated at each point before summing.}
\label{fig:fit}
\end{figure*}

\subsection{Spectral Fitting} \label{sec:spectral fitting}

Spectral classification of pre-main sequence stars has traditionally been performed using empirical spectral templates from libraries such as \citet{1998Pickles}, \citet{2009Rayner}, or \citet{2017Manara,2013Manara}. These templates, derived from observations of well-characterized stars, allow for direct comparison with observed spectra. The classification is often based on the relative depths of absorption features which are sensitive to $T_{\mathrm{eff}}$, such as TiO and VO bands in M-type stars \citep{1991Kirkpatrick}, or atomic absorption lines, such as the hydrogen Balmer series and HeI and FeI lines in early-type stars (B, A, and F) \citep{2001Gray}. However, for low-metallicity environments, no comprehensive empirical spectral library exists for pre-main sequence stars, making direct comparison with template spectra not possible.

Given this limitation, we employed theoretical model spectra from the BT-Settl AGSS 2008 grid \citep{2011Allard}, which provide synthetic spectra across a range of effective temperatures \(T_{\text{eff}}\), surface gravities ($ {log~g }$), and metallicities ($ {Z}$). The spectral fitting was performed over the continuum on wavelength range of 5500 to 9300 \AA\ to maximize sensitivity to key absorption features while avoiding noisy regions in the spectra with strong telluric contamination. Although the MUSE spectra cover 4500 to 9500 \AA, fitting tests across different wavelength ranges indicated that the chosen interval provided the most reliable fits.

The fitting process involved the following parameters:

\begin{itemize}

\item\textbf{Effective Temperature (\(T_{\text{eff}}\)):} The original BT-Settl model grid samples \(T_{\text{eff}}\) between 2600 and 50000\,K in steps of 100\,K. To achieve a finer resolution, we interpolated the model spectra to intermediate temperatures using the radial basis function interpolation method from \texttt{SciPy}, thereby expanding the temperature grid to a step size of 10\,K.

\item \textbf{Metallicity (\(Z\)):} We adopted a cluster metallicity of \(Z = -0.7\) ($Z = 0.2 \, Z_\odot$ corresponds to $\log(0.2) \approx -0.7$), based on our analysis described in Section~\ref{sec:metallicity} and consistent with literature values from \citet{Negueruela2015}. Since BT-Settl models are available only at discrete metallicities, we interpolated linearly in flux space between the \(Z = -1.0\) and \(Z = -0.5\) model spectra (at the same \(T_{\text{eff}}\) and \(\log g\)) to generate synthetic spectra at \(Z = -0.7\).

\item \textbf{Surface Gravity ($\log g$):} Surface gravity was varied over the range $\log g = 3.0$–$4.5$. Most pre-main-sequence stars were best fit with $\log g \approx 3.5$, consistent with typical values for young, low-mass stars.

\item \textbf{Visual Extinction ($A_V$):} The extinction was varied from 0 to 5 mag, with an initial estimate based on the reported cluster extinction of $A_V \sim 2.5$--3.0 from \citet{Guarcello2021} and \citet{Kalari2015}.

\item \textbf{Total-to-Selective Extinction Ratio ($R_V$):} The extinction law was assumed to follow a standard $R_V = 3.1$, consistent with the value derived from unevolved main-sequence stars in the region \citep{Delgado2010}.

\item\textbf{Veiling ($r$):} We adopted a wavelength-dependent veiling model, following the approach of \citet{2021Fang}. The accretion continuum was modeled as a blackbody spectrum with a fixed temperature of 7000~K.
The total flux at each wavelength was computed as:

\[
F_\lambda^{\mathrm{veiled}} = F_\lambda^{\mathrm{model}} + r \cdot F_\lambda^{\mathrm{bb}},
\]

where \(F_\lambda^{\mathrm{model}}\) is the flux from the stellar photosphere from BT-Settl models, \(F_\lambda^{\mathrm{bb}}\) is the normalized blackbody spectrum at 7000~K, and \(r\) is the veiling factor. The veiling factor is varied from 0 to 2 in steps of 0.02.

\end{itemize}

The fitting procedure consisted of several steps. First, the synthetic model spectra were convolved with a Gaussian kernel and resampled onto the observed wavelength grid to match the spectral resolution of MUSE. The wavelength-dependent blackbody component was then added to account for veiling, followed by extinction correction using the \citet{2023Gordon} extinction law. The resulting model spectra were normalized at 7465~\AA\ and compared to the normalized observed spectra of each object over the same wavelength range. The best-fit parameters were obtained by minimizing the reduced \(\chi^2\) over a finely sampled grid of effective temperature (\(T_{\text{eff}}\)), visual extinction (\(A_V\)), \(\log g\))  and veiling factors. Strong emission lines and noisy wavelength regions were masked to prevent artificial inflation of the \(\chi^2\) values. The final parameter set corresponds to the minimum \(\chi^2\) solution and was further validated through visual inspection of the residuals and flux ratios.

Bolometric luminosities were estimated by first dereddening the observed spectra using the best-fit \(A_V\) and then subtracting the veiling contribution according to the best-fit veiling factor. The best-fit synthetic model was scaled down to match the flux level of the veiling- and extinction-corrected observed spectrum, and its flux was integrated over all wavelengths to derive the total luminosity. Uncertainties in luminosity were calculated by propagating the errors in \(T_{\text{eff}}\) and \(A_V\).

Fit quality was assessed through multiple diagnostics, including visual inspection of the best-fit spectrum, a Hess diagram of the \(\chi^2\) distribution, and the median absolute deviation between the observed and model spectra (see Figure~\ref{fig:fit}). To validate the continuum-based fitting, a spectral index method was used to estimate \(T_{\text{eff}}\) for K–M type stars. Both methods show excellent agreement, with a median offset of 25~K and a dispersion of \(\pm200\)~K ( see Appendix ~\ref{appendix:spectral_index} for details).

\subsection{Spectroscopic Identification of PMS Stars}

Pre-main-sequence stars in our sample were primarily identified through spectroscopic diagnostics, including accretion signatures and features indicative of stellar youth such as Li absorption and Balmer emission. This approach was necessitated by the limited ancillary photometric coverage in Dolidze~25, particularly in the infrared, which prevented the construction of color–color or color–magnitude diagrams and reliable spectral energy distributions (SEDs). The main spectroscopic features used for classification are described below.

\paragraph{Balmer Emission:}

Balmer emission lines are ubiquitous in our sample and are widely used as diagnostics of accretion in young stellar objects. The wavelength range of MUSE covers the H$\alpha$ line at 6563~\AA\ and the H$\beta$ line at 4861~\AA. These lines originate primarily from warm ionized gas in the accretion flow, with additional contributions from hydrogen recombination in the post-shock region \citep{1998Muzerolle, 1998Calvet, 2012Alencar}, and are considered reliable tracers of mass accretion \citep{Hartmann2016}. In our sample, 30\% of the sources exhibit H$\alpha$ equivalent widths greater than 10~\AA\ and H$\beta$ EWs above 5~\AA, consistent with ongoing accretion. Whereas, 15\% of the sources show both lines in absorption, characteristic of hotter stars with weak or no accretion signatures.

\paragraph{Lithium Absorption:}

The Li\,\textsc{i}~$\lambda6708$\,\AA\ absorption line serves as a crucial indicator of youth for stars with spectral types later than G5. During pre-main sequence evolution, lithium is rapidly depleted through convective mixing in low-mass stars. Therefore, its presence typically signifies stellar ages younger than $\sim$20\,Myr, particularly for stars with masses below $\sim$1\,M$_\odot$ \citep[e.g.,][]{2016Bouvier,2020Olney,2023Jeffries}. About 40\% of sources in our sample show absorption in Li I with EW $>0.25$~\AA. 

\paragraph{Other emission lines: He~I, Ca~II , Na~I :}

The He~I \(\lambda 5876\)~\AA\ emission line, though weak in our spectra, are also associated with accretion shocks \citep{2001Beristain}. In our sample, the He~I lines are detected in emission predominantly in the spectra with strong H\(\alpha\) emission. However, only 3\% of sources show He\,\textsc{i} emission with EW $< -1$~\AA, suggesting a smaller fraction with strong accretion-driven ionization.

The Ca~II infrared triplet (IRT) at $\lambda\lambda\,8498$, 8542, 8662\,\AA\ and the Na\,\textsc{i} doublet at $\lambda\lambda\,8183$, 8195\,\AA\ are commonly used as diagnostics of accretion and surface gravity, respectively, in young stars. However, both features are detected in fewer than 1\% of the sources in our sample. Only one source exhibits all three Ca\,\textsc{ii} lines in emission, four sources show two lines, and ten display at least one line with equivalent width $\mathrm{EW} < -1$\,\AA. The reduced line strength of Ca\,\textsc{ii} emission is consistent with inner disk metal depletion, as reported by \citet{Marbely2023,Marbely2024} in transitional disks. Their studies demonstrate that reduced gas-phase abundances of refractory elements such as calcium in the inner disk can lead to significantly weaker Ca\,\textsc{ii} emission, consistent with our observations.

\subsection{Cluster Membership} \label{sec:membership}

We identify members of Dolidze 25 by combining astrometric information from \textit{Gaia} DR3 \citep{2023Gaia} with youth indicators derived from optical spectroscopy. This approach enables us to construct a robust census of pre-main-sequence members across a range of stellar masses and evolutionary stages, extending sensitivity to sources not accessible through \textit{Gaia} due to extinction or faintness.

\subsubsection{Astrometric Selection from \textit{Gaia}}

Astrometric membership was assessed using \textit{Gaia}~DR3 proper motions and parallaxes (see Section~\ref{sec:distance}). This analysis identified 26 astrometric members within the MUSE field of view, the majority of which are bright (\(L > 1\,L_\odot\), \(r_{\mathrm{mag}} < 22\)) and likely relatively massive stars. Among these, approximately 50\% show Balmer lines (H$\alpha$, H$\beta$) in absorption, consistent with their classification as early-type stars.

Only 9 sources exhibit at least one spectroscopic youth indicator (e.g., lithium absorption or accretion-related emission), suggesting their PMS status. The remaining stars lack prominent Balmer absorption or emission and do not exhibit lithium features, showing mostly featureless continuum spectra. These may be weak-lined pre-main-sequence stars with faint or absent disk emission, indicating little to no active accretion, or stars with warm photospheres where lithium absorption is not detectable (see Figure~\ref{fig:cmd}).

\subsubsection{Spectroscopic Membership via Youth Diagnostics}

For fainter sources not detected in \textit{Gaia} (\(r_{\mathrm{mag}} > 22\)), we relied on optical spectral features indicative of stellar youth. These include lithium absorption and permitted emission lines associated with accretion. Additionally, X-ray detections were used to identify young stellar objects lacking reliable astrometric and spectroscopic information.

A source was considered a probable cluster member if it satisfies any of the following conditions:

\begin{itemize}

    \item \textit{Lithium absorption:} Presence of the Li\,\textsc{i} 6708\,\AA{} line with an equivalent width (EW) $>$ 0.25\,\AA{}. This threshold effectively distinguishes PMS stars from older field populations, as lithium is depleted on timescales $\lesssim$20\,Myr in low-mass stars \citep{2016Bouvier}.
    
    \item \textit{Accretion Indicators:} Evidence for ongoing accretion was identified through emission in at least two of the following lines, with each exhibiting an equivalent width greater than 1\,\AA\ in emission (i.e., EW $< -1$\,\AA) to ensure the features are clearly distinguishable from the continuum: H$\alpha$ (6563\,\AA), H$\beta$ (4861\,\AA), He\,\textsc{i} (5876\,\AA), Ca\,\textsc{ii} IRT (8542\,\AA), or Na\,\textsc{i} (8190\,\AA).
    
    \item \textit{X-ray detection:} Significant detection in at least one energy band in the \textit{Chandra} X-ray catalog of \citet{Guarcello2021}. Low-mass stars (K0--M5) dominate the X-ray detections due to their strong magnetically driven coronal activity. Weak-lined T Tauri stars, which often lack strong infrared excess from disks, are especially well-detected, as their X-ray emission is not significantly attenuated by circumstellar material. X-ray detections are particularly effective in identifying diskless or weakly accreting members that may be missed in infrared or optical disk-emission based selection methods.

\end{itemize}

\begin{figure*}
\centering
\includegraphics[width=0.99\textwidth]{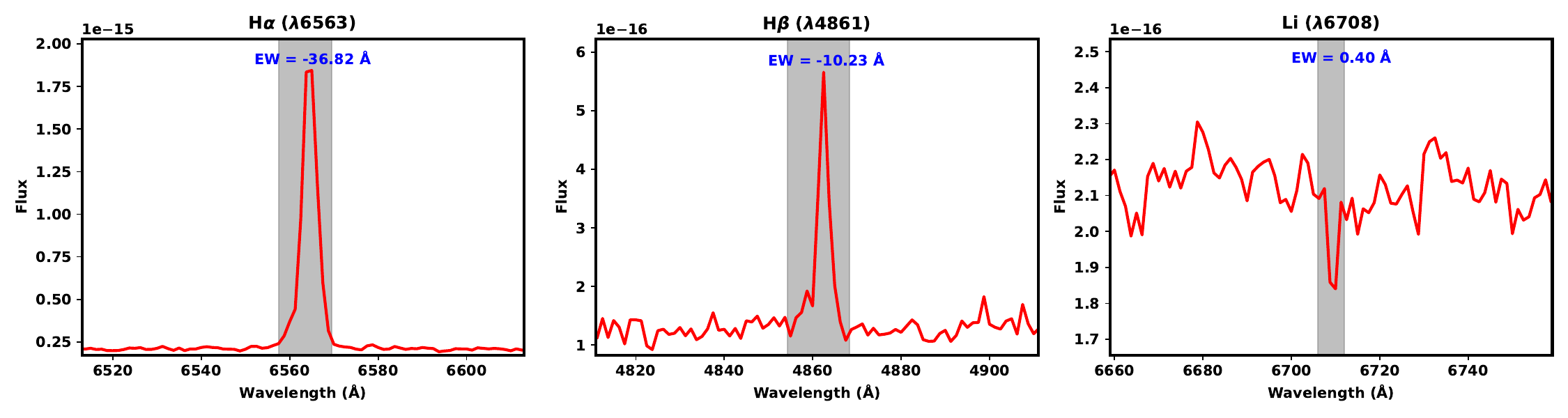} 
\caption{H$\alpha$, H$\beta$, and Li\,\textsc{i} line profiles of a typical accreting star with $T_{\mathrm{eff}} \sim 3500$\,K. The H$\alpha$ (6563\,\AA) and H$\beta$ (4861\,\AA) lines are seen in emission, while the Li\,\textsc{i} (6708\,\AA) line appears in absorption with corresponding equivalent widths indicated in each panel.}
\label{fig:hahbli}
\end{figure*}

\begin{figure}
\centering
\includegraphics[scale = 0.35]{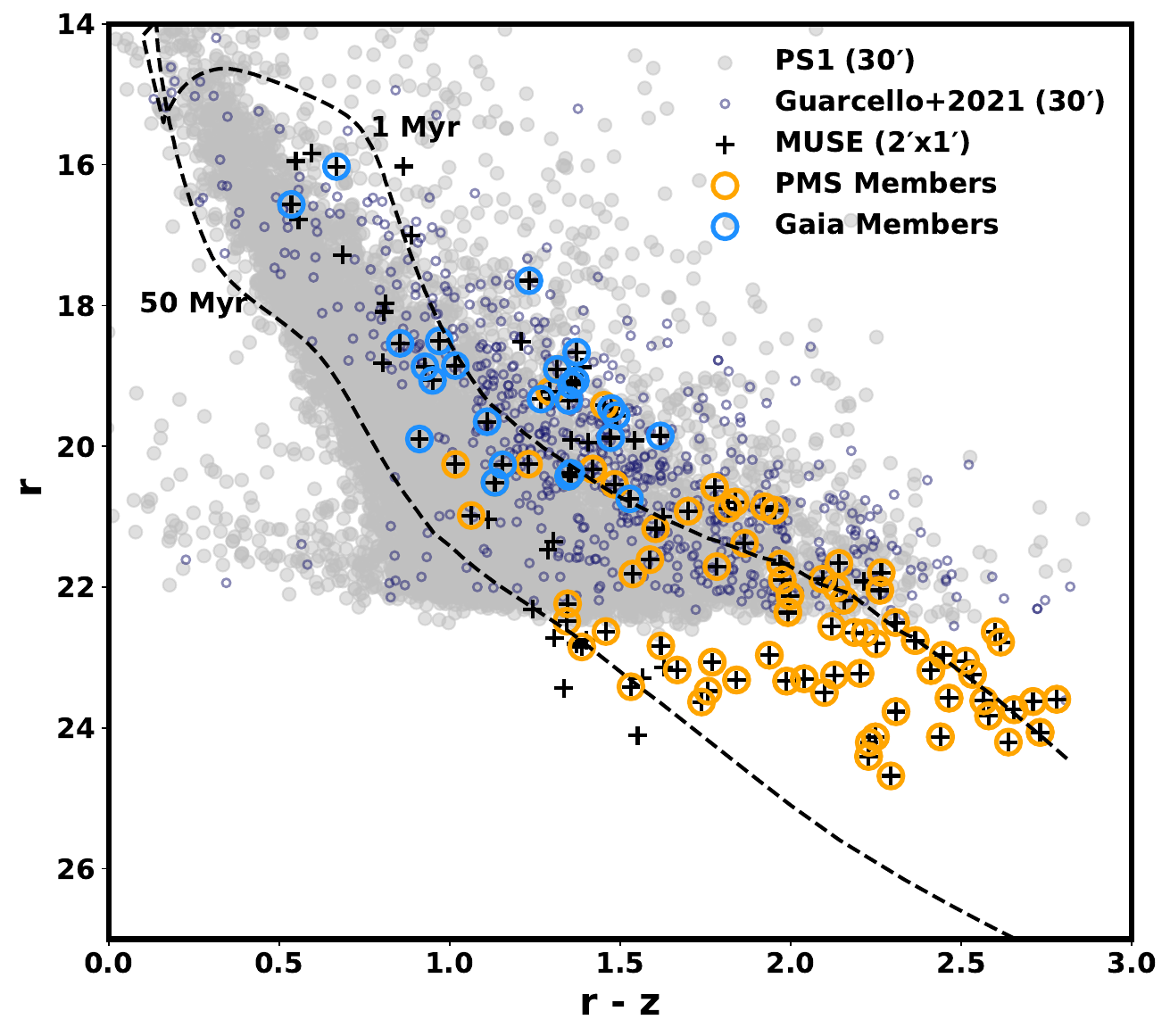} 
\caption{Color-magnitude diagram (\(r\) vs. \(r - z\)) constructed using synthetic photometry extracted from MUSE spectra. Grey points represent all detections from Pan-STARRS within a 30\arcmin\ radius centered on the cluster. X-ray and infrared-selected members from \citet{Guarcello2021}, covering the same 30\arcmin\ region, are indicated by navy blue circles. Black crosses mark all sources detected by MUSE with S/N \(> 5\) within its \(2 \times 1~\mathrm{arcmin}^2\) field of view. Probable pre-main-sequence members identified via MUSE spectroscopy are highlighted with orange circles, while \textit{Gaia}-based members are shown as blue circles. For reference, the 1 Myr and 50 Myr PARSEC isochrones, corrected for an extinction of \(A_V = 2\) mag and distance of 4.2 kpc, is overplotted. The depth of the MUSE data enables the detection of members down to \(r \sim 25\), significantly fainter than the Pan-STARRS photometric limit of \(r \sim 22\).}
\label{fig:cmd}
\end{figure}

\begin{figure}
\centering
\includegraphics[scale = 0.35]{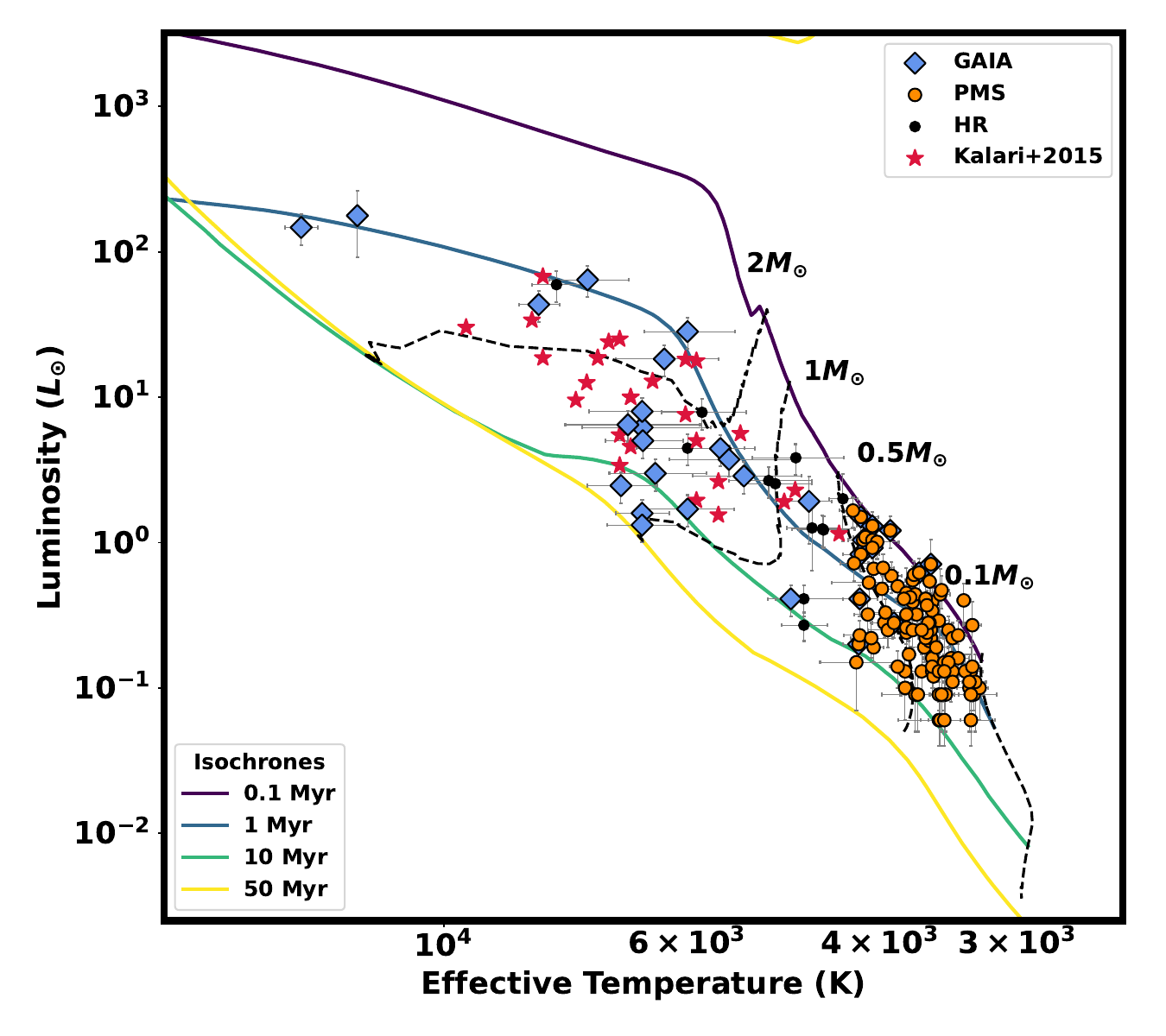} 
\caption{HR diagram of the studied stellar population. Orange circles represent spectroscopically identified PMS members, blue diamonds indicate GAIA-based members, and black circles denote members identified solely by their HR diagram positions. Red stars mark sources previously studied spectroscopically by \citet{Kalari2015,2011Cusano}. Solid lines represent \textsc{Parsec} isochrones for ages 0.1 to 50 Myr, and dotted lines show evolutionary tracks for indicated masses. PMS sources are primarily low-mass stars ($< 1 \, M_{\odot}$) with ages ranging from 0.1 to 10 Myr and a median age of 1.5 Myr, consistent with YSO characteristics.}
\label{fig:hr_diagram}
\end{figure}

All sources satisfying at least one of the above criteria were placed on the Hertzsprung–Russell (HR) diagram using effective temperatures and extinction-corrected luminosities derived from spectral fitting (see Section~\ref{sec:spectral fitting}). All of the spectroscopically identified sources are located above the 10\,Myr PARSEC isochrone, with inferred masses below $1\,M_\odot$, occupying regions characteristic of pre-main-sequence stars (see Figure~\ref{fig:hr_diagram}). Hence, based on the presence of spectroscopic youth or accretion indicators, together with their positions in the HR diagram, we identify 95 PMS members among 132 sources with reliable spectral types. These objects are predominantly located in the low-mass regime ($M \lesssim 1\,M_\odot$) and exhibit varying levels of accretion activity.

Among these, 48 sources show Li\,\textsc{i}~6708\,\AA\ absorption with equivalent widths $> 0.25$\,\AA, 65 show accretion- or activity-related emission lines, and 14 are detected in X-rays. These diagnostics show substantial overlap: 35 sources exhibit both Li absorption and emission lines, 6 display Li absorption along with X-ray detection, and another 6 show emission lines and X-ray detection. Only 3 sources exhibit all three youth indicators simultaneously. Among the X-ray detected sources, 9 (64\% of the X-ray members) are confirmed by at least one additional indicator. A few additional PMS candidates are identified solely through X-ray emission, likely including WTTS, which are X-ray bright despite weak or absent spectroscopic features of accretion, as well as sources where veiling or low S/N hinders the detection of spectroscopic youth/accretion indicators.

In summary, our final membership census within the central region of Dolidze~25, spanning an area of \(2 \times 1~\mathrm{arcmin}^2\) (corresponding to \(\sim 3~\mathrm{pc}^2\) at a distance of 4.2~kpc), identifies a total of 95 pre-main-sequence members based on spectroscopic youth diagnostics. An additional 26 Gaia-selected stars exhibit astrometric coherence with the cluster but lack definitive youth signatures in our spectra. Furthermore, 11 sources are not classified as members via Gaia kinematics or PMS accretion/youth indicators; however, their positions on the HR diagram suggest probable membership. These sources, characterized by effective temperatures above 4000~K, likely represent more evolved cluster members that are undetected in Gaia and no longer show prominent accretion or youth-related emission features. Combining these, we confirm a total of 132 cluster members in Dolidze~25 with stellar parameters determined, of which 95 are classified as PMS sources. The parameters for all confirmed members are provided in Table~\ref{table: all_members}.

Given the large distance to Dolidze~25, our spectroscopic sample is likely incomplete at the lowest stellar masses. In particular, we do not identify any cluster members with effective temperatures below 3100\,K. This likely reflects a detection bias, as such sources are intrinsically faint and their spectra, when observed, tend to exhibit low signal-to-noise ratios, resulting in unreliable spectral fits. Incompleteness may also affect higher-temperature sources at the faint end of the luminosity distribution, such as older or less massive stars with low intrinsic luminosities. Furthermore, the limited spatial coverage of the MUSE observations implies that our sample underrepresents the full population of low-luminosity pre-main-sequence stars in the cluster.

Figure~\ref{fig:cmd} shows the \(r\) vs. \(r - z\) color–magnitude diagram constructed from synthetic photometry derived from MUSE spectra for all sources with S/N \(> 5\). The diagram shows the detection limit of MUSE compared to existing datasets: Pan-STARRS reaches a limiting magnitude of \(r \sim 22\), whereas MUSE spectroscopy allows identification of members as faint as \(r \sim 25\). Very few faint members ($r > 22$) are recovered by the \citet{Guarcello2021} catalog, which covers a larger 30\arcmin\ region but is likely limited by photometric sensitivity at fainter magnitudes.

\section{Stellar Properties and Cluster Characteristics} \label{sec:results}

The physical properties of the confirmed members-including effective temperature, extinction, and luminosity-were obtained from spectral model fitting, as described in Section~\ref{sec:spectral fitting}. The median visual extinction ($A_V$) derived for the region from the best fits of the members is 1.9 $\pm0.9$ magnitudes, consistent with previous studies ($A_V = 2.4$~mag; \citealt{Delgado2010}). The effective temperatures (\(T_{\text{eff}}\)) and bolometric luminosities (\(L\)) were used to construct the HR diagram and to derive stellar parameters such as mass, age and accretion rates. 

\subsection{Stellar Ages, Masses, and Radii}

To characterize the stellar population, we constructed the HR Diagram using spectroscopically derived effective temperatures (\(T_{\text{eff}}\)) and bolometric luminosities (\(L\)) (see Fig.~\ref{fig:hr_diagram}). The HRD serves as a key diagnostic for assessing the evolutionary state of the stars, allowing mass and age estimates through comparison with theoretical models. We employed \textsc{Parsec} v1.2S isochrones and evolutionary tracks \citep{2012Bressan}, computed for a metallicity of \(Z = 0.2\, Z_\odot\), and spanning ages from 0.1 to 50 Myr. 

Stellar masses and ages were derived by minimizing the distance between each source’s position on the HRD and the isochrone grid. Uncertainties were estimated via 1000 Monte Carlo realizations, wherein \(T_{\mathrm{eff}}\) and \(\log L\) were randomly perturbed within their respective Gaussian errors. Each realization was matched to the closest point on the isochrone to infer mass and age, and the 16th and 84th percentiles of the resulting distributions were adopted as the 1\(\sigma\) confidence intervals.

Stellar radii (\(R\)) were calculated using the Stefan–Boltzmann relation between luminosity, effective temperature, and radius. This approach assumes negligible residual interstellar extinction, which is justified, as the bolometric luminosities were derived from dereddened model spectra despite the average extinction in the region being \(A_V \sim 2\) mag.

\begin{figure*}
\centering
\includegraphics[width=0.99\textwidth]{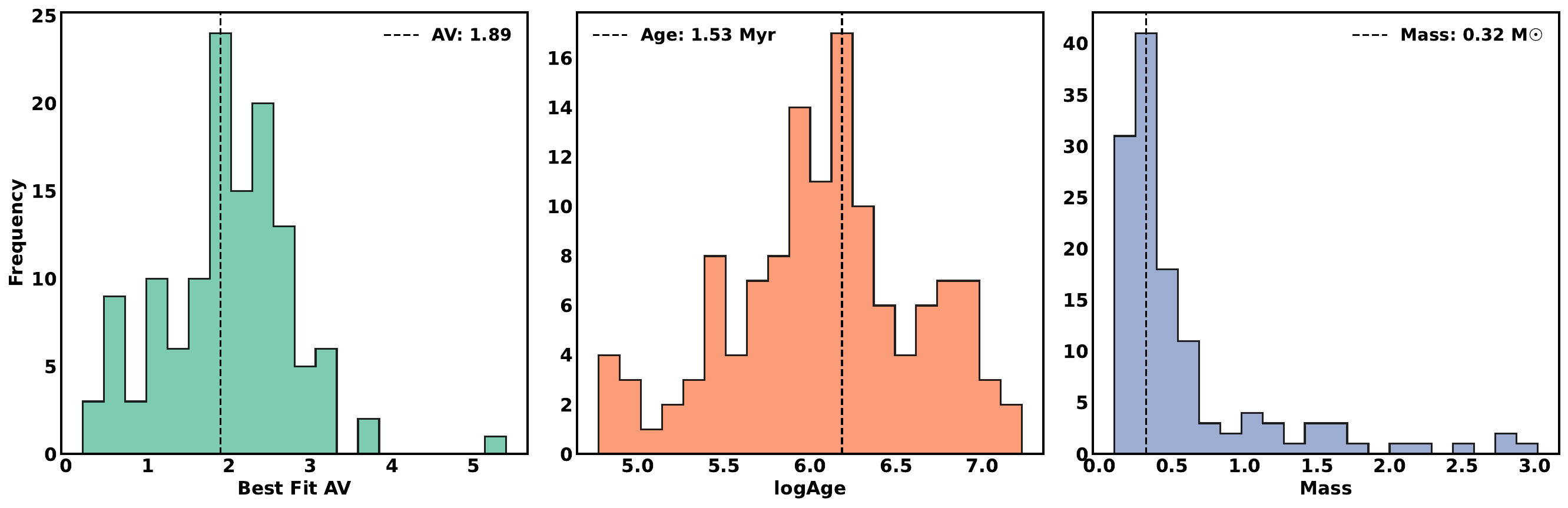} 
\caption{Distribution of best-fit extinction (A\textsubscript{V}), age, and mass for identified cluster members derived from spectral fitting and \textsc{Parsec} isochrones. The peak values for each parameter are annotated for reference.} 
\label{fig:age_hist}
\end{figure*}

The median age of the cluster is found to be $\sim 1.5$~Myr, with individual stellar ages for most sources ranging from 0.1 to 10~Myr. This age distribution confirms that the stars in our sample are in the early stages of stellar evolution, during which significant accretion activity is still ongoing. The majority of sources have estimated masses between 0.1 and 1~$M_\odot$, with a mass distribution peaking at $0.32~M_\odot$, similar to that observed in other young clusters \citep{2021Damian,2024Gupta}. However, given the incompleteness of the current sample, no firm conclusions can be drawn regarding the characteristic mass and IMF (see Figure~\ref{fig:age_hist} for the age and mass distributions).

A comparison of the HR diagram (Figure 9) with the color-magnitude diagram (Figure 8) reveals a notable discrepancy in the inferred ages of the PMS population. While the CMD shows a population of PMS stars scattered towards the 50~Myr isochrone, the same sources are consistently placed at much younger ages ($<10$~Myr) in the HRD. This discrepancy is primarily caused by the combined effects of accretion-related continuum excess and extinction on optical colours. Veiling substantially modifies $r$-band magnitudes and $r-i$ colours, causing accreting sources appear both brighter and bluer in CMDs, thereby mimicking older ages \citep{2025Piscarreta}. Furthermore, intrinsic luminosity spreads and variable extinction can contribute to this scatter \citep{2014Soderblom}. Our spectroscopic analysis directly accounts for the veiling continuum and provides individual extinction estimates, resulting in the tighter and systematically younger age distribution observed in the HRD. This comparison underscores that spectroscopically derived HRDs yield more reliable age estimates than CMDs.

\subsection{Accretion rate }

\begin{figure*}
\centering
\includegraphics[width=0.9\textwidth]{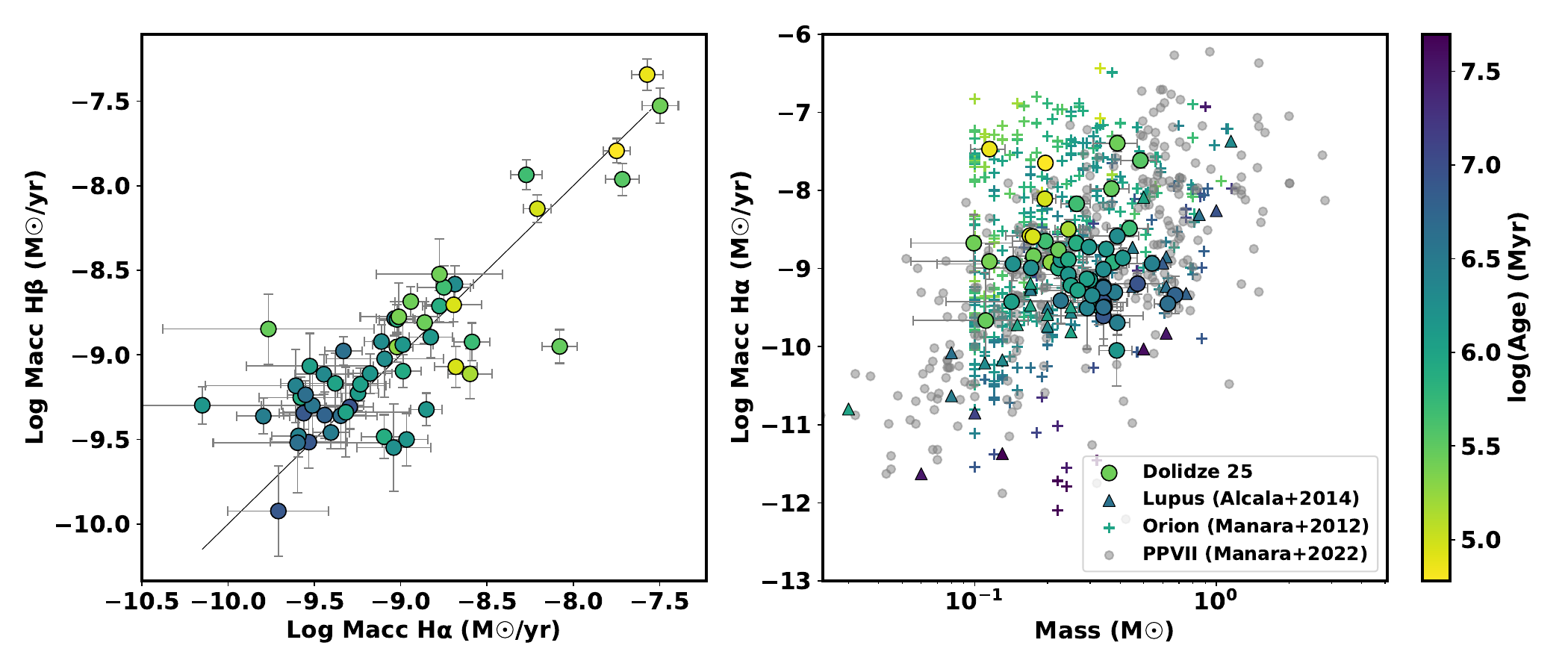} 
\caption{\textbf{Left:} Mass accretion rate derived from H$\alpha$ versus H$\beta$. Both are in good agreement with each other. \textbf{Right:} Mass accretion rate (\(\log \dot{M}_{\text{acc}}\)) as a function of stellar mass for pre-main-sequence stars in Dolidze~25 (this work), compared with similarly aged (1--3 Myr) clusters: Lupus \citep{2014Alcala} and Orion \citep{2012Manara}. Gray points represent mass accretion rates of young stellar objects in nearby solar-metallicity star-forming regions within 300\,pc, including Lupus, Orion, Taurus, Chamaeleon~I and II, Ophiuchus, Corona Australis, and Upper Scorpius, compiled by \citet{Manara2022}. Dolidze~25 sources exhibit mass accretion rates comparable to those in solar-metallicity environments, despite the lower metallicity of the cluster.}
\label{fig:acchahb}
\end{figure*}

The identification of H$\alpha$ emitters is based on their equivalent width measurements and line profile characteristics, as discussed in \citet{2023Ashraf}. Accretion rates are calculated for sources with H$\alpha$ equivalent widths greater than 10~\AA, based on the H$\alpha$ luminosity derived from the line's equivalent width and the interpolated continuum flux at the H$\alpha$ wavelength. We applied the empirical relationship to convert H$\alpha$ luminosity to accretion luminosity using the values $a = 1.13$ and $b = 1.74$ \citep{Alcala2017}. The accretion rate was then derived using the formula:
\[\dot{M}_{\text{acc}} = \frac{L_{\text{acc}} R_{\text{star}}}{0.8 G M_{\text{star}}}\]
where $G$ is the gravitational constant, $R_{\text{star}}$ is the stellar radius derived from the Stefan-Boltzmann law, and $M_{\text{star}}$ is the stellar mass obtained from isochrone fitting. The factor of 0.8 arises from the assumption that the accretion flows originate at a magnetospheric radius of approximately \( R_{\rm m} \sim 5\,R_\star \) \citep{1994Shu}. The mass accretion rates estimated from H$\alpha$ span a range from \(10^{-8}\) to \(10^{-10}\,M_\odot\,\text{yr}^{-1}\), with a median value of \(8 \times 10^{-10}\,M_\odot\,\text{yr}^{-1}\).

The H$\beta$ emission line was detected in most sources exhibiting H$\alpha$ emission with EW $< -10$\,\AA. We derived the accretion rates using H$\beta$ equivalent widths, following the method outlined above and employing the empirical coefficients from \citet{Alcala2017}. The accretion rates derived from H$\beta$ are in good agreement with those derived from H$\alpha$ with a standard deviation of 0.3. 

To ensure consistency with literature values, we adopted accretion rates derived from the H$\alpha$ line for subsequent analysis. As shown in Figure~\ref{fig:acchahb} (right panel), the mass accretion rates of young stellar objects in Dolidze~25 are broadly consistent with those observed in nearby solar-metallicity star-forming regions within 300\,pc, including Lupus, Orion, Taurus, Chamaeleon~I and II, Ophiuchus, Corona Australis, and Upper Scorpius \citep{Manara2022}. The reference sample, shown as gray points in the figure, is compiled in Table~1 of \citet{Manara2022} and primarily based on homogeneous accretion diagnostics using VLT/X-Shooter spectra. Despite the significantly lower metallicity of Dolidze~25 (approximately 1/5 of the solar value), its accretion rates do not show any notable systematic deviation from those of similar-mass and similar-age stars in these solar-metallicity environments. This suggests that metallicity may not be the dominant factor controlling the mass accretion rate at the observed ages and masses.

This finding aligns with the results of \citet{Kalari2015}, who estimated mass accretion rates for pre-main-sequence stars with masses between 0.5 and 2.5~$M_\odot$ in the broad region Sh~2-284, including Dolidze~25, using optical spectroscopy from the SALT telescope. Despite the low metallicity, their analysis found no significant difference in accretion rates compared to solar-metallicity PMS stars in the same mass range. We find consistent results in the low mass regime: despite the significant difference in metallicity, the accretion processes appear to be relatively unaffected. The physical parameters of the accreting sources including mass, age, and accretion rates are presented in Appendix~\ref{table:accretors}.

\section{Discussion} \label{sec:discussion}

\begin{figure*}
\centering
\includegraphics[width=0.9\textwidth]{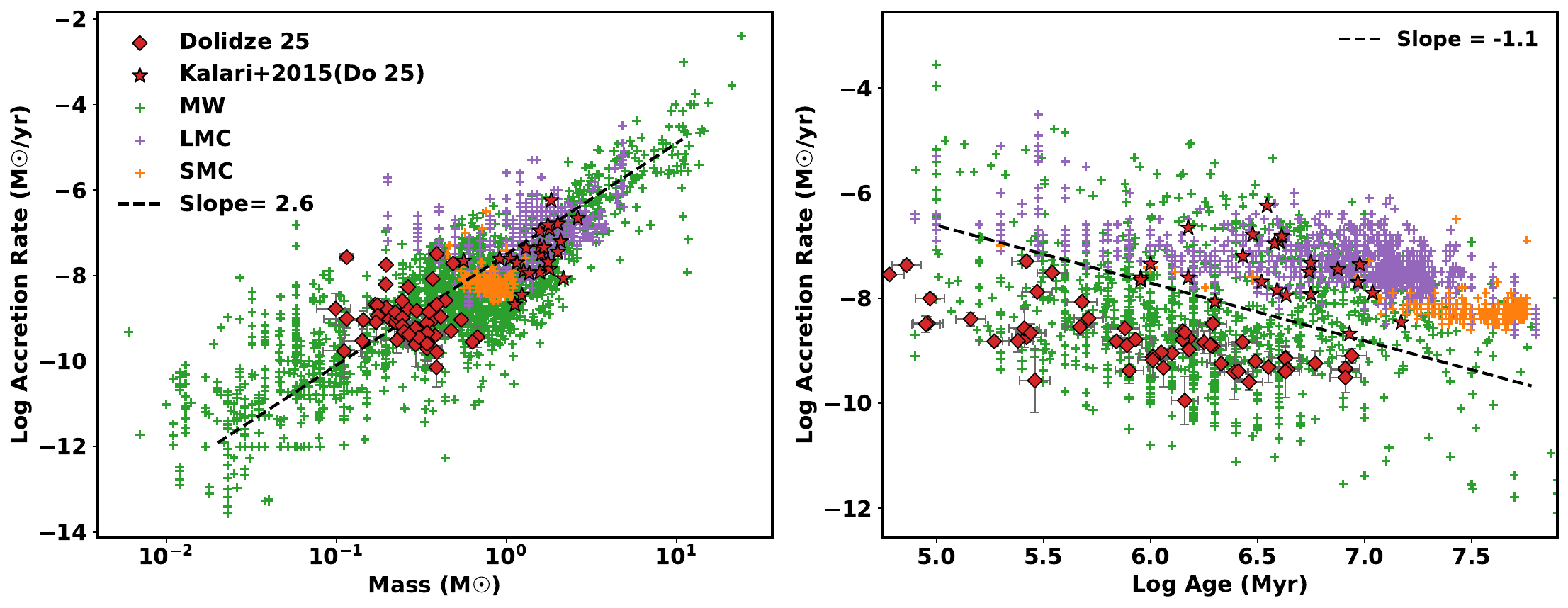} 
\caption{Stellar mass vs. mass accretion rate (left) and stellar age vs. mass accretion rate (right) relations for pre-main-sequence stars in Dolidze~25, compared with literature samples from the Milky Way (MW)—Lupus \citep{2014Alcala}, Orion \citep{2012Manara}, M8 \citep{2024Venuti}, IC~1396 \citep{2011Barentsen}, \citet{2015Fairlamb}, \citet{2023Nidhi}, \citet{2024Damian}, \citet{2025Ben}; the Large Magellanic Cloud (LMC)—\citet{2012Spezzi, 2019Biazzo,2017DeMarchi, 2022Carini, 2004Romaniello}; and the Small Magellanic Cloud (SMC)—\citet{2023Vlasblom, 2011DeMarchi, 2013Demarchi, 2023Tsilia}. Mass accretion rates for sources in Dolidze~25 with masses \(\sim0.5 - 2\,M_\odot\), determined from SALT spectra by \citet{Kalari2015}, are also included. The dashed lines indicate the best-fit relations derived exclusively from the compiled Milky Way data, with the associated slopes annotated.} 

\label{fig:acc_lmc_smc}
\end{figure*}

\subsection{The Role of Metallicity and Environment in Regulating Mass Accretion}

The influence of metallicity on mass accretion rates are important in understanding the evolution of circumstellar disks around pre-main sequence stars, with significant implications for stellar growth and planet formation. Observations of exoplanet populations show a strong positive correlation between stellar metallicity and the likelihood of hosting giant planets \citep[e.g.,][]{Fischer2005,Fulton2021,2021Zhu}. Planet formation efficiency increases with metallicity, as a higher dust-to-gas ratio enhances grain growth and facilitates planetesimal formation \citep[e.g.,][]{Miotello2022,2024Matsukoba}. This interplay between metallicity, accretion, and disk lifetime is therefore central to understanding both stellar growth and the conditions that set the efficiency and outcomes of planet formation across different environments.

Theoretically, lower metallicity environments are predicted to exhibit higher protostellar accretion rates due to differences in disk thermodynamics and opacity \citep{2009Hosokawa, 2014Tanaka, 2015Machida, 2018Nakatani}. In such environments, the reduced metal content leads to a lower dust-to-gas ratio and reduced opacity, resulting in less efficient radiative cooling. Consequently, the disk gas remains hotter, enhancing its scale height and promoting more rapid infall onto the central protostar. Additionally, the decreased dust content reduces shielding from far-ultraviolet radiation, rendering the disk more susceptible to photoevaporation. As a result, disk dispersal is expected to proceed more rapidly, shortening the disk lifetime in metal-poor regions \citep{2018Nakatani}.

To assess whether the low metallicity of Dolidze~25 affects protostellar accretion in practice, we examined how its accretion properties compare with those in other Galactic and extra galactic regions. The dependence of mass accretion rate (\(\dot{M}_{\text{acc}}\)) on stellar mass (\(M_*\)) is well-characterized by a power-law relationship of the form \(\dot{M}_{\text{acc}} \propto M_*^\alpha\), where the exponent \(\alpha\) typically ranges between 1.5 and 3.1 \citep[e.g.,][]{2003Muzerolle, 2004Calvet, 2008Herczeg, 2009Fang, 2015Manara, Alcala2017, Hartmann2016}. We compiled pre-main sequence accretion rates from literature sources, including \citet{2012Manara, 2015Fairlamb, Alcala2017, 2018Vioque, 2023Nidhi, 2024Venuti, 2024Damian, 2025Ben}, covering multiple Galactic star-forming regions. These data collectively follow a scaling relation of \(\dot{M}_{\text{acc}} \propto M_*^{2.6}\), with a large scatter. The sources in Dolidze~25 align well with this relation and expected scatter, showing no evidence for elevated accretion rates relative to other Galactic populations (Figure~\ref{fig:acc_lmc_smc}).

The relation between accretion rate and stellar age also follows a power-law decline, typically expressed as \(\dot{M}_{\text{acc}} \propto t^{\alpha}\), where \(\alpha\) ranges from \(-1.6\) to \(-1.0\) \citep{2012Manara, 2014Antoniucci, 2014Venuti, Hartmann2016}. Our analysis of this age-accretion relationship, presented in the right panel of Figure~\ref{fig:acc_lmc_smc}, yields a best-fit slope of \(-1.1\), consistent with previous studies. The plot shows substantial scatter, largely driven by the underlying mass distribution, the lower-mass sources (\(M \sim 0.3\,M_\odot\)) from this study are located at the lower part of the distribution, while the higher-mass (\(\gtrsim 1\,M_\odot\)) sources from \citet{Kalari2015} exhibit higher accretion rates.

These results demonstrate that the mass and age dependence of mass accretion rates are consistent across galactic environments studied to date. The factor of five lower metallicity does not lead to any significant changes in accretion rates.

In contrast to our results, differences in accretion rates in the LMC and SMC had previously been attributed to low metallicity \citep{2012Spezzi, 2019Biazzo, 2023Vlasblom}. Although those accretion rates of YSOs follow a similar dependence on mass as seen in the Milky Way, the accretion rates persist for tens of Myr. Maintaining such high accretion rates over long timescales would require some drastic differences in envelope and disk evolution for the SMC and LMC. The differences are unlikely explained by metallicity alone.
Beyond metallicity, other environmental factors such as ambient gas density, radiation fields, and stellar clustering can also significantly influence accretion efficiency (eg. \citealt{2024Damian}). The central O7-type star in Dolidze~25 is a source of ultraviolet radiation (Lyman continuum photon rate,  \(\log Q_{\mathrm{HI}} \sim 48.5\)), which could lead to external photoevaporation of nearby protoplanetary disks. This process can truncate the disk radius and reduce its mass reservoir, leading to a decrease in the mass accretion rate over time \citep{2017Rosotti, 2023Mauco}. However, such harsh radiative environments are not unique to Dolidze~25; similar or even more extreme conditions exist in the LMC and SMC, which host rich populations of O and B-type stars with intense UV radiation fields \citep{2014Sabin,2022ARickard}. Furthermore, the influence of external photoevaporation is expected to be most prominent within the inner $\sim$0.5 pc of massive stars of similar spectral type (eg. \citealt{2023Mauco}), implying that any external impact in Dolidze~25 is likely confined to the immediate vicinity of its central O-type star. At larger cluster radii, where most of our targets are located, the radiation field strength declines rapidly, and the observed accretion rates being comparable to those in other Galactic star-forming regions indicate that external photoevaporation is not the dominant factor determining the observed accretion rates in this cluster(see Figure \ref{fig:r_vs_ha} in Appendix)

Instead, other environmental factors, such as ambient gas density may play a more significant role. Recent studies have found correlations between higher accretion rates and local gas density \citep{2024Winter, 2025Rogersb}, consistent with the idea that large, dense clusters like those in the LMC/SMC could exhibit systematically different accretion properties despite similar metallicities. Dolidze~25 is relatively sparse with a lower stellar and gas density, which may help explain why the accretion rates are broadly similar to typical Galactic regions, highlighting that metallicity alone may not be the primary driver.
An alternative explanation of the SMC and LMC results are systematic uncertainties in stellar properties. Our results (see Section \ref{sec:spectral fitting}) indicate that the stellar photospheres are similar enough between low- and solar metallicity YSOs to not introduce large systematic uncertainties in relative fits. However, a possible explanation may be that stellar ages may be systematically old when accreting YSO SEDs are fit with stellar photospheres \citep{2025Piscarreta}.

Interpreting these findings also requires consideration of observational biases. Accretion rates in the LMC and SMC are inferred from photometric excesses, and the sampled populations are biased toward more massive stars ($M_* > 1\,M_\odot$). Future spectroscopic measurements of accretion rates in these regions would help to mitigate such biases and provide more direct, reliable comparisons across different environments \citep{2024Demarchi_1, 2025Biazzo}. Moreover, the limited sample size in Dolidze~25 due to the restricted spatial coverage of a $1\times2$~arcmin central region near an O-type star is not enough to conclude on the accretion behavior of the larger cluster. Expanding the spatial coverage of this region and incorporating additional diagnostics, such as UV excess measurements, would be essential for obtaining a more comprehensive understanding of accretion behavior and its environmental dependencies. 

In summary, although lower metallicity environments are theoretically expected to sustain higher accretion rates, our analysis of Dolidze~25 reveals no indication of enhanced accretion compared to solar-metallicity regions. Other environmental factors may reduce or offset the effects of metallicity, leading to accretion rates similar to those observed in solar-metallicity regions. Future studies using larger and more diverse samples will be essential to better understand how metallicity and environment together shape accretion processes.

\begin{figure*}
\centering
\includegraphics[width=0.9\textwidth]{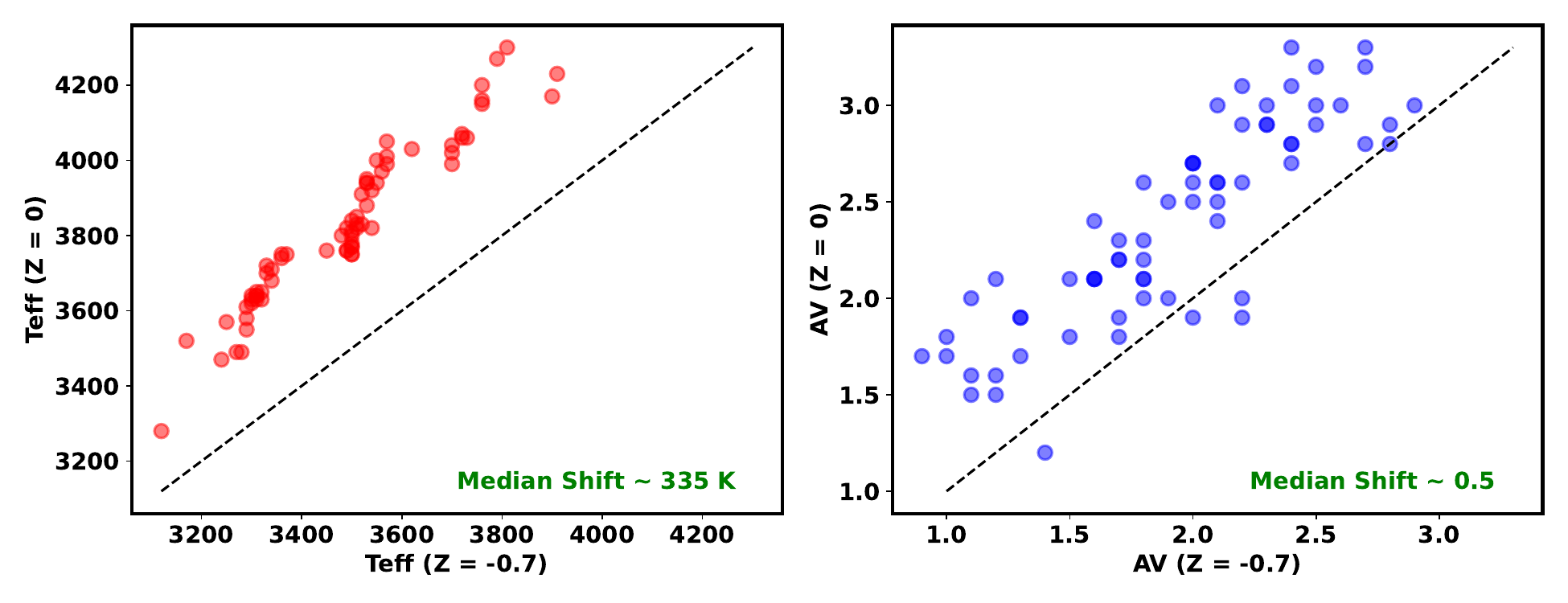} 
\caption{Difference in effective temperature (\(T_{\text{eff}}\)) and extinction (Av) estimates derived using solar metallicity templates compared to low metallicity templates for pre-main sequence sources. The dashed line indicates the 1:1 relation for reference. The results show a systematic overestimation of both \(T_{\text{eff}}\) and Av when solar metallicity templates are used.}
\label{fig:solar scatter}
\end{figure*}

\begin{figure*}
\centering
\includegraphics[width=0.9\textwidth]{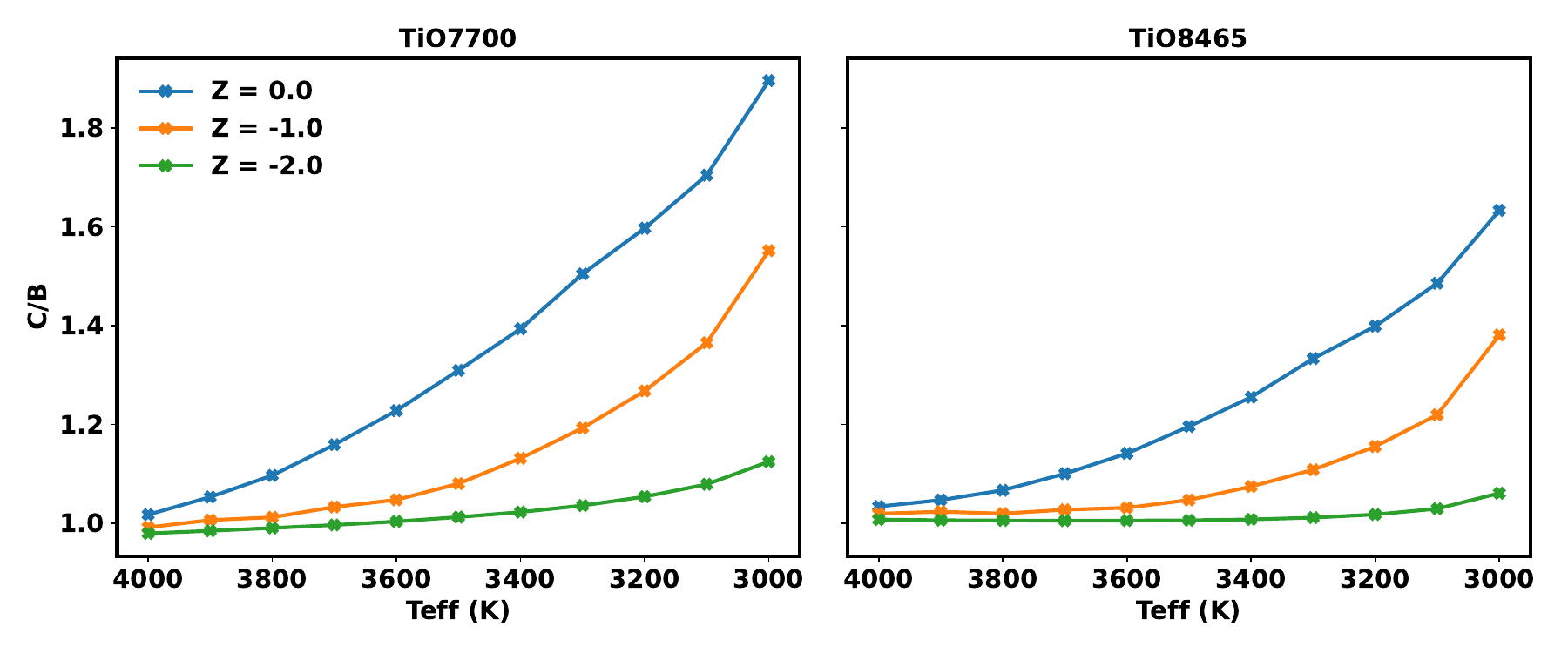} 
\caption{Variation of the spectral index as a function of effective temperature (\(T_{\text{eff}}\)) for BT-Settl models at different metallicities. The curves demonstrate a systematic shift in the spectral index vs temperature relation with respect to metallicity.}

\label{fig:siindex_vs_metallicity}
\end{figure*}

\subsection{How Accurate Are Solar Metallicity Models for Fitting Low Metallicity Stars?}

Due to the lack of low-metallicity observational templates in the literature, most previous studies have relied on solar metallicity templates to fit low-metallicity spectra. To investigate the impact of this approach, we conducted a comparative analysis by fitting our low-metallicity MUSE spectra of pre-main-sequence members using both solar and subsolar metallicity models from the BT-Settl grid. The fitting methodology and parameter ranges were kept consistent with those described in Section \ref{sec:spectral fitting} of our study.

Our analysis revealed that the best-fit solutions, as indicated by the minimum chi-squared (\(\chi^2\)) values, were consistently obtained when using low-metallicity templates, confirming a better match to the observed spectral features. In contrast, fits using solar metallicity models systematically overestimated key physical parameters—most notably the effective temperature (\(T_{\text{eff}}\)) and extinction (\(A_V\)). This trend was robust across both the continuum-fitting and spectral index approaches. While our primary focus has been on low-mass stars, we note that similar trends in fitting behavior persist at higher effective temperatures (\(T_{\text{eff}} > 5000\,\text{K}\)). As shown in Fig.~\ref{fig:solar scatter}, the derived \(T_{\text{eff}}\) values from solar metallicity models were uniformly higher than those obtained using low-metallicity models.

Spectral indices measured from model spectra show that specific molecular features—particularly TiO bands—are sensitive to metallicity at fixed \(T_{\text{eff}}\). The indices exhibit systematically lower values in low-metallicity sources, indicating weaker TiO absorption (see Figure~\ref{fig:siindex_vs_metallicity}). For instance, an observed TiO index of 1.3 is matched at \(T_{\text{eff}} \approx 3200\,\text{K}\) using a low-metallicity template (\(Z = -0.5\)), whereas a solar-metallicity model (\(Z = 0\)) reproduces the same index only at a significantly higher \(T_{\text{eff}} \approx 3500\,\text{K}\). This implies that when solar-metallicity templates are used to fit low-metallicity stars, the derived effective temperatures are systematically overestimated. This systematic offset arises because solar-metallicity models predict stronger TiO absorption at a given temperature, and the fitting process compensates for the mismatch by assigning a higher temperature to reproduce the observed, weaker molecular features. Therefore, using metallicity-inappropriate models leads to biased parameter estimates and misrepresents the true temperature scale of low-metallicity stars.

These findings emphasize the importance of using metallicity-appropriate models for accurate spectral fitting. Relying on solar metallicity templates can introduce systematic biases in derived parameters, which may propagate into misinterpretations of stellar properties and evolutionary states in metal-poor environments. To help address this gap, we provide a set of flux-calibrated and extinction-corrected spectra for low-mass PMS stars in a metal-poor environment, derived from our VLT/MUSE observations. These spectra constitute a valuable resource for refining spectral classification and parameter estimation in low-metallicity star-forming regions, and contribute toward building a more representative template library for future studies.

\section{Conclusion} \label{sec:conclusion}

We conducted a comprehensive spectroscopic study of the low-mass pre-main-sequence stellar population in the low-metallicity star-forming region Dolidze 25 (\(Z \sim 0.2 \, Z_\odot\)) using VLT-MUSE observations. Below, we summarize the methodologies employed and the key findings derived from this analysis:

\begin{itemize}
    
    \item VLT-MUSE optical spectroscopy (4750-9350\,\AA) was used to observe the central \(2'\times1'\) region of Dolidze~25, identifying 132 cluster members- 95 based on spectroscopic youth indicators, 26 from \textit{Gaia} astrometry and 11 from HR diagram positions.
    
    \item Source characterization was performed through spectral fitting using low-metallicity BT-Settl models, yielding effective temperatures, extinctions, and luminosities. Stellar masses (median 0.3 \(M_\odot\)) and a median age of \(\sim 1.5\) Myr were estimated using PARSEC isochrones, confirming a young, low-mass population.
    
    \item Mass accretion rates, derived from H$\alpha$ emission, range from \(10^{-10}\) to \(10^{-8}\,M_\odot\,\text{yr}^{-1}\), with a median of \(8 \times 10^{-10}\,M_\odot\,\text{yr}^{-1}\). These values are comparable to those in solar-metallicity regions such as Lupus and Orion. No systematic enhancement or suppression in accretion rate is observed, despite the metallicity of Dolidze~25 being approximately one-fifth of solar.
    
    \item Use of solar-metallicity templates results in systematic overestimations of effective temperature (\(T_{\text{eff}}\)) and extinction (\(A_V\)), emphasizing the necessity of low-metallicity models for accurate parameter estimation in metal-poor environments. We provide a set of of low-metallicity low mass spectra for future studies.

\end{itemize}

This work demonstrates the capability of MUSE to probe low-mass stellar populations in distant and metal-poor star-forming regions. The observed accretion properties provide new constraints on disk evolution in sub-solar metallicity environments, indicating that metallicity may subtly influence disk dispersal timescales and accretion dynamics. Future studies extending spatial coverage and incorporating longer wavelength data will be essential to further dissect the metallicity–accretion relationship in diverse galactic environments.

\begin{acknowledgments}

We thank the anonymous referee for their constructive comments that improved the quality of this manuscript. This work is based on observations collected at the European Southern Observatory under ESO programme 098.C-0435(A). It is also based in part on data obtained from the Pan-STARRS1 Surveys (PS1). This research has made use of the VizieR catalogue access tool, CDS, Strasbourg, France (DOI: \href{https://doi.org/10.26093/cds/vizier}{10.26093/cds/vizier}). JJ acknowledge the DST-SERB, Gov. of India, for the POWER grant (No: SPG/2021/003850). CFM is funded by the European Union (ERC, WANDA, 101039452). Views and opinions expressed are however those of the author(s) only and do not necessarily reflect those of the European Union or the European Research Council Executive Agency. Neither the European Union nor the granting authority can be held responsible for them. We acknowledge the use of ChatGPT and GitHub Copilot for assistance with language refinement and coding support during the preparation of this work.

\end{acknowledgments}

\clearpage

\nocite{2009Gordon,2019Fitzpatrick,2022Decleir}
\bibliography{ref}{}
\bibliographystyle{aasjournal}

\clearpage

\appendix

\section{Scaled nebular background subtraction}\label{app:nii_subtraction}

The nebular emission in Dolidze 25 is highly inhomogeneous, influenced by the ionized gas environment near the central O7 type star (see H$\alpha$ intensity map Figure \ref{fig:ha_map}). To quantify the degree of spatial variability, we extracted spectra from multiple sky regions devoid of stars across the MUSE field using 4-pixel apertures. The equivalent widths of nebular emission lines show a range of values: H$\alpha$ exhibited a median EW of approximately $-70$\,\AA, with values ranging from $-10$ to $-90$\,\AA; [N\,\textsc{ii}]~$\lambda6583$ had a median EW of $-10$\,\AA, spanning $-1$ to $-12$\,\AA; and [S\,\textsc{ii}]~$\lambda6731$ showed a median EW of $-5$\,\AA, with a range from 0 to $-6$\,\AA. This broad variation reflects the uneven structure of the ionized gas within the H\,\textsc{ii} region. 

The direct subtraction of nebuar flux from the source often resulted in either residual nebular contamination or unphysical negative fluxes due to oversubtraction. To mitigate these effects, we implemented a scaled background subtraction method following the approach of \citet{2024DeMarchi,2024Rogers}, who utilized the strength of He\,\textsc{i} 1.869 $\micron$ line to scale nebular spectra in NGC 3603. In our study, the [N\,\textsc{ii}] $\lambda6583$ and [S\,\textsc{ii}] $\lambda6731$ forbidden lines were adopted as scaling tracers to correct for local fluctuations in nebular emission.

The selection of these lines as scaling metrics rests on the assumption that their emission originates predominantly from the nebular background, with negligible stellar contributions. To validate this assumption, we examined high-resolution X-shooter spectra of K- and M-type young stellar objects from the $\sigma$ Orionis, Lupus, and Taurus regions \citep{2017Manara,2013Manara}. These spectra exhibited no detectable [N\,\textsc{ii}] or [S\,\textsc{ii}] emission, with equivalent widths close to zero, confirming their utility as nebular tracers for low-mass pre-main-sequence stars.

The subtraction procedure involved an iterative optimization process. An initial scaling factor of 0.01 was applied to the nebular spectrum, which was then subtracted from the source spectrum. The sum of the absolute EWs of the [N\,\textsc{ii}] and [S\,\textsc{ii}] lines in the residual spectrum was computed, and the scaling factor was incrementally adjusted in steps of 0.01 until the absolute sum of the EWs was minimized. The optimal scaling factor was selected based on two criteria: (1) the EWs of the reference lines were driven as close to zero as possible, and (2) the number of negative flux values in the corrected spectrum was minimized.

\begin{figure}
\centering
\includegraphics[width=0.6\textwidth]{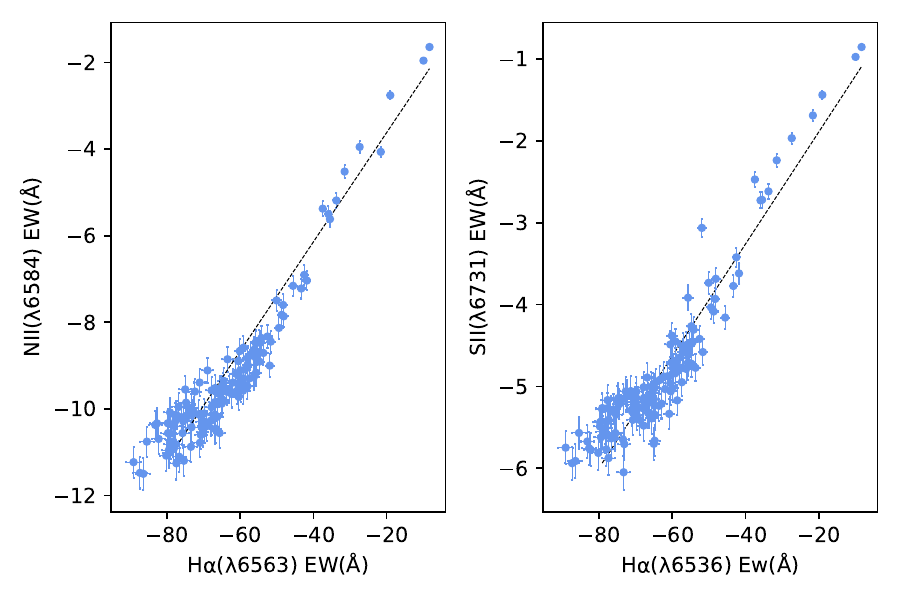}
\caption{Scatter plot of [N II] (red) and [S II] (blue) equivalent widths versus H-alpha equivalent width, showing a strong positive correlation. The standard deviations are 0.61 and 0.36 for NII and SII respectively.}
\label{fig:nii_sii_halpha}
\end{figure}

The entire nebular spectrum was scaled using the optimal scaling factor and subsequently subtracted from the source spectrum. This background scaling assumes that subtracting [N\,\textsc{ii}] and [S\,\textsc{ii}] emission also proportionally removes the nebular Balmer emission—particularly H$\alpha$, which is the primary accretion diagnostic. This assumption is supported by the observed correlation between the fluxes of [N\,\textsc{ii}], [S\,\textsc{ii}], and H$\alpha$ in nebular-dominated regions across the MUSE field. A direct comparison of their EWs, illustrated in Figure~\ref{fig:nii_sii_halpha}, demonstrates a tight linear relationship, with standard deviations 0.61 and 0.36 for NII and SII respectively, justifying the use of these forbidden lines as proxies for nebular H$\alpha$ subtraction. 

\begin{figure}
\centering
\includegraphics[width=0.9\textwidth]{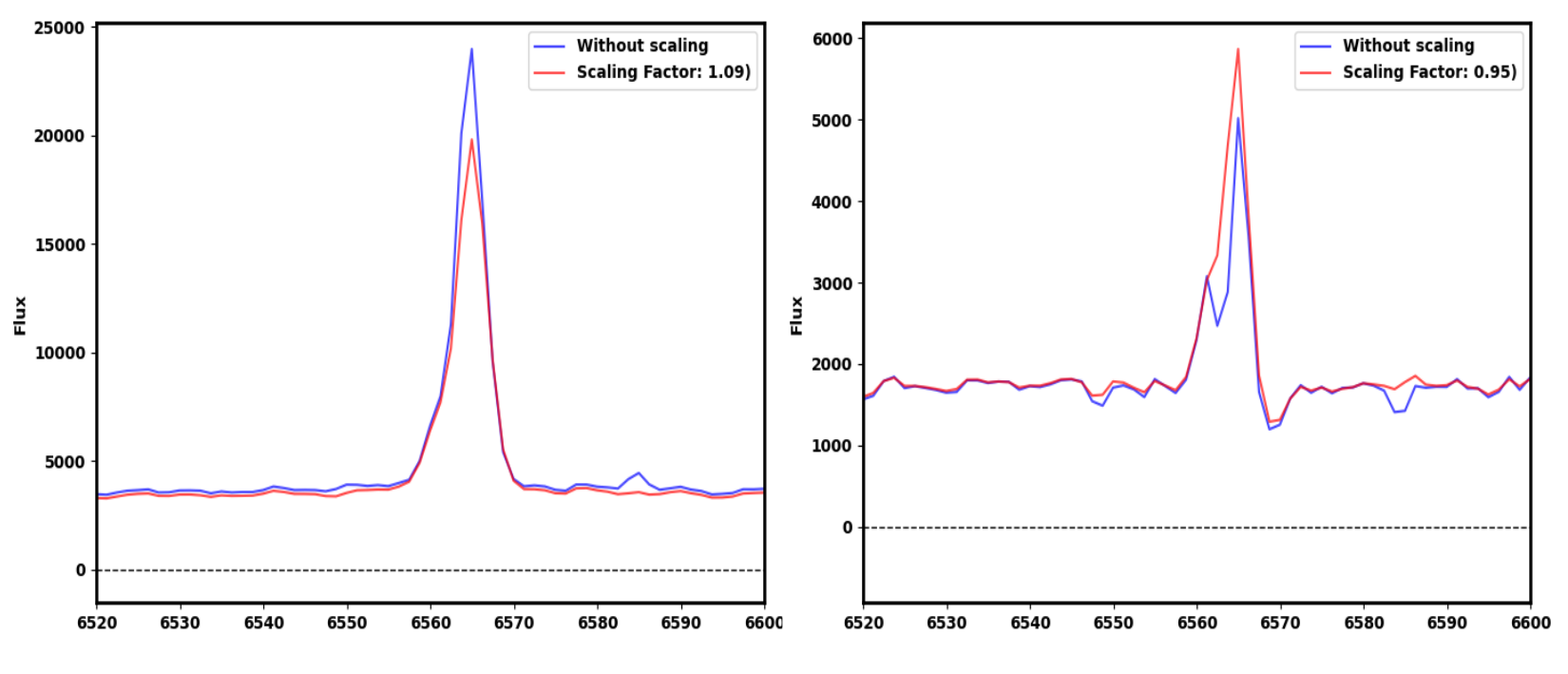}
\caption{Examples of nebular background subtraction using the [N\,\textsc{ii}]-based scaling method. \textbf{Left:} A case of slight under-subtraction or a weak intrinsic [N\,\textsc{ii}] emission from the star; this may lead to a mild underestimation of the accretion rate (up to $\sim$10\%). \textbf{Right:} A typical case of oversubtraction from unscaled methods. The scaled subtraction yields a cleaner H$\alpha$ profile and reduces the occurrence of unreal negative fluxes in the spectrum.}
\label{fig:scaled}
\end{figure}

Figure~\ref{fig:scaled} shows two example background subtractions. The left panel illustrates a case of slight under-subtraction or a weak intrinsic [N\,\textsc{ii}] emission from the star itself. In the case of intrinsic [N\,\textsc{ii}] emission, the background scaling may lead to oversubtraction of H$\alpha$, potentially resulting in an underestimation of the accretion rate by up to $\sim$10\%. This value should be regarded as an upper limit, since tests with reduced equivalent widths (5–10~\AA) showed that the actual impact on accretion rates is smaller, with changes in H$\alpha$ EW typically less than 3~\AA. The right panel depicts an example of oversubtraction, the most common case in our sample. Visual inspection of the extracted spectra, guided by the [N\,\textsc{ii}] $\lambda\lambda6549,6583$ lines and the absence of artificial negative fluxes, ensured robust subtraction of the nebular component. The resulting line-scaled background subtraction produced cleaner H$\alpha$ profiles and significantly reduced unrealistic negative flux occurrences. The scaling factor is less than one for 95\% of the sources, with a median value of $0.95 \pm 0.2$, indicating that circumstellar contributions to [N\,\textsc{ii}] and [S\,\textsc{ii}] emission are minimal.

\section{Spectral Index-Based Temperature Estimation} \label{appendix:spectral_index}

To complement and cross-verify the continuum-based fitting method, a spectral index-based approach was also employed to estimate the effective temperatures (\(T_{\text{eff}}\)) of K–M type stars. This technique relies on the temperature-sensitive TiO molecular absorption bands that dominate the optical spectra of late-type stars and vary predictably with \(T_{\text{eff}}\). These bands correspond to distinct vibrational transitions of TiO, with their relative strengths and depths varying systematically with \(T_{\text{eff}}\) due to changes in molecular dissociation and excitation conditions in stellar atmospheres, making them reliable tracers of photospheric temperature in cool stars \citep{2014Herczeg, 2014Damiani}. 

\begin{figure}
\centering
\includegraphics[width=0.99\textwidth]{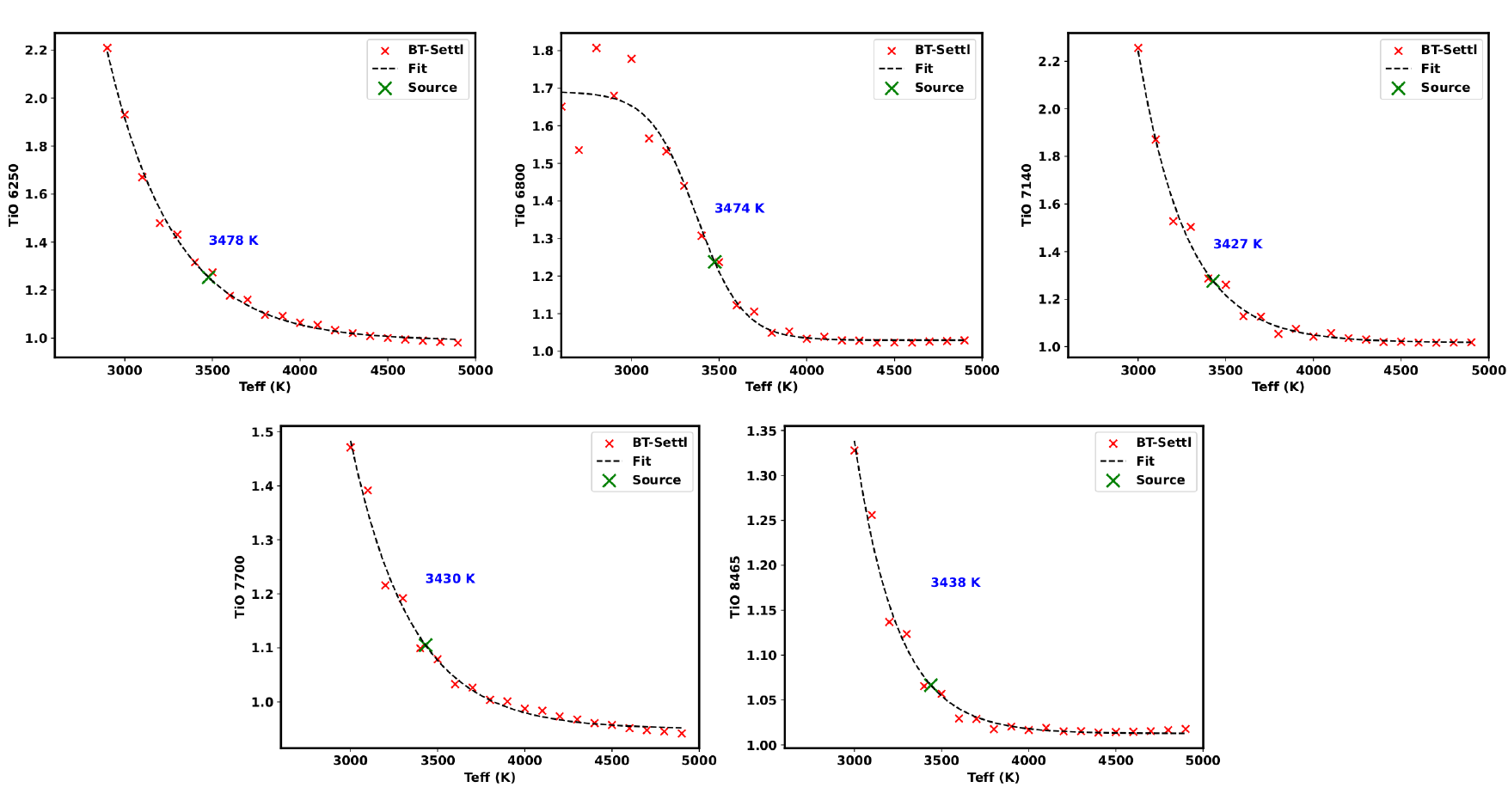} 
\caption{Spectral index versus \(T_{\text{eff}}\) for 5 TiO features centered at 6250 \AA, 6800 \AA, 7140 \AA, 7700 \AA, and 8465 \AA. Each subplot represents the relationship between spectral index and \(T_{\text{eff}}\) based on BT-Settl model spectra with $Z = -0.7 $ and  $log~g = 4.0 $. Red crosses indicate spectral indices computed from the models, while the green cross represents the observed spectral index from an example MUSE source. The corresponding \(T_{\text{eff}}\) values derived from the observed spectral index are annotated in blue in each subplot. The final effective temperature (\(T_{\text{eff}}\)) of the source is adopted as the mean of the five individual estimates, yielding \(T_{\text{eff}} = 3450 \pm 25\,\text{K}\), where the uncertainty is given by the standard deviation of the measurements.}
\label{fig:si fit}
\end{figure}

A grid of synthetic spectra from the BT-Settl atmospheric models \citep{2011Allard} was used to establish a \(T_{\text{eff}}\)- Spectral Index calibration curve . The models span \(T_{\text{eff}} = 2500-5000\) K, metallicity \(Z = -0.7\), and surface gravity \(\log g = 4.0\), appropriate for metal-poor K-M type stars. Each synthetic spectrum was convolved to the resolution of our observed data, and the five TiO indices, centered at 6250~\AA, 6800~\AA, 7140~\AA, 7700~\AA, and 8465~\AA, were computed following the definitions in \citet{2014Herczeg}, using the median flux within the specified continuum and bandpass ranges for each feature. The indices were then plotted against \(T_{\text{eff}}\) and a fourth-order polynomial function was fitted to the data to derive a continuous calibration curve ( see Figure~\ref{fig:si fit}). This approach accounts for the nonlinear response of TiO absorption strength to temperature, particularly in the critical 2500–5000 K range where TiO bands emerge and saturate.
 
For each observed source, the same five spectral indices were measured from its flux-calibrated, telluric-corrected MUSE spectrum. The resulting indices were independently translated into \(T_{\text{eff}}\) using the model-based calibration curves. The final \(T_{\text{eff}}\) for each object was computed as the mean of the five individual values, while their standard deviation gives the error.

\begin{figure}
\centering
\includegraphics[scale = 0.2]{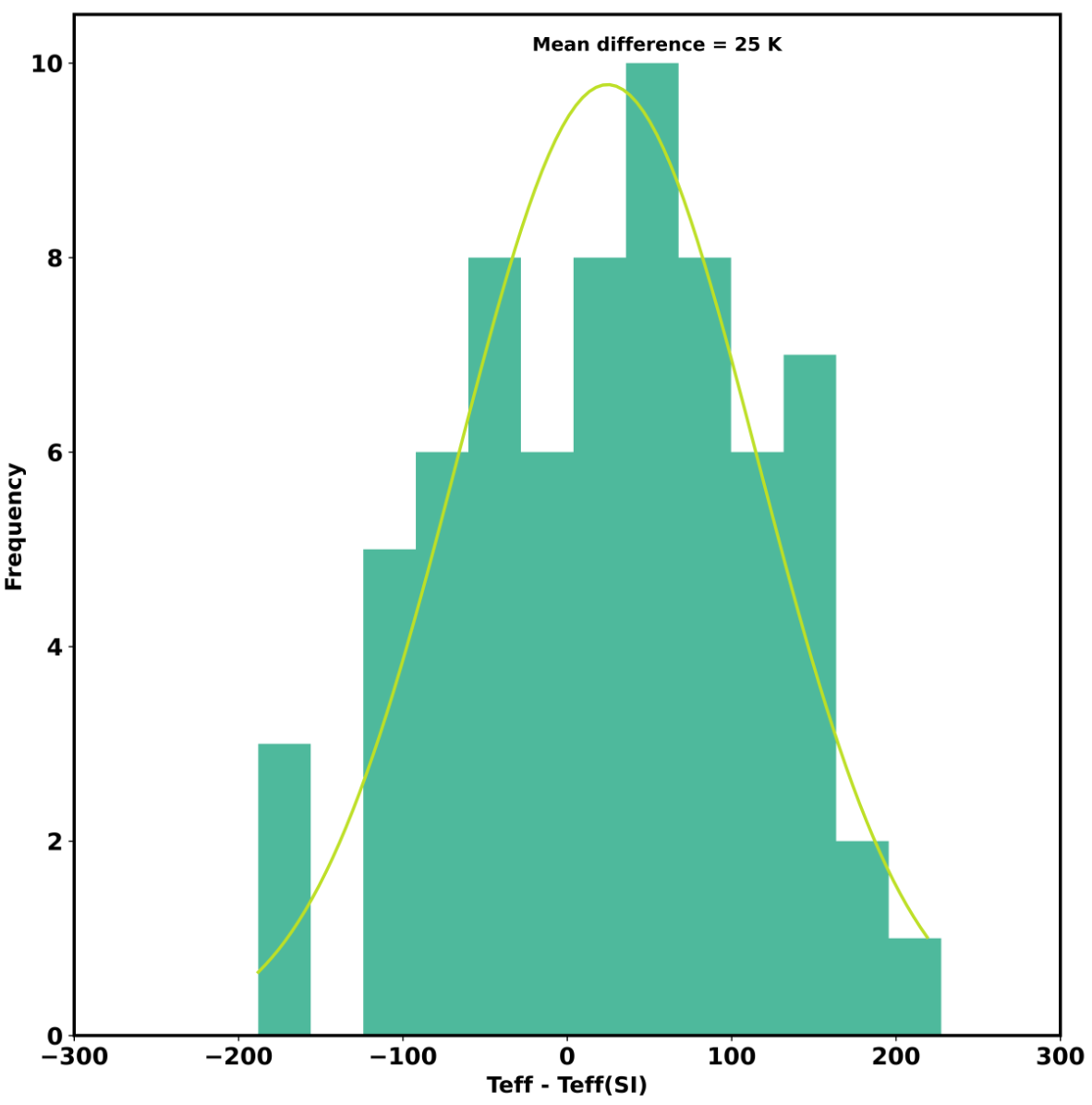} 
\caption{A histogram of the difference between the best-fit $T\textsubscript{eff}$ values obtained from the two methods (i.e., \(\chi^2\) minimization between the source and template fluxes and \(\chi^2\), from Spectral Indices vs $T\textsubscript{eff}$ relation for K-M type stars derived from BT-Settl templates. The distribution exhibits a median offset of 25\,K, with differences in $T_\mathrm{eff}$ reaching up to $\pm 200\,\mathrm{K}$.}
\label{fig:si histogram}
\end{figure}

\begin{figure}
    \centering
    \includegraphics[width=0.8\textwidth]{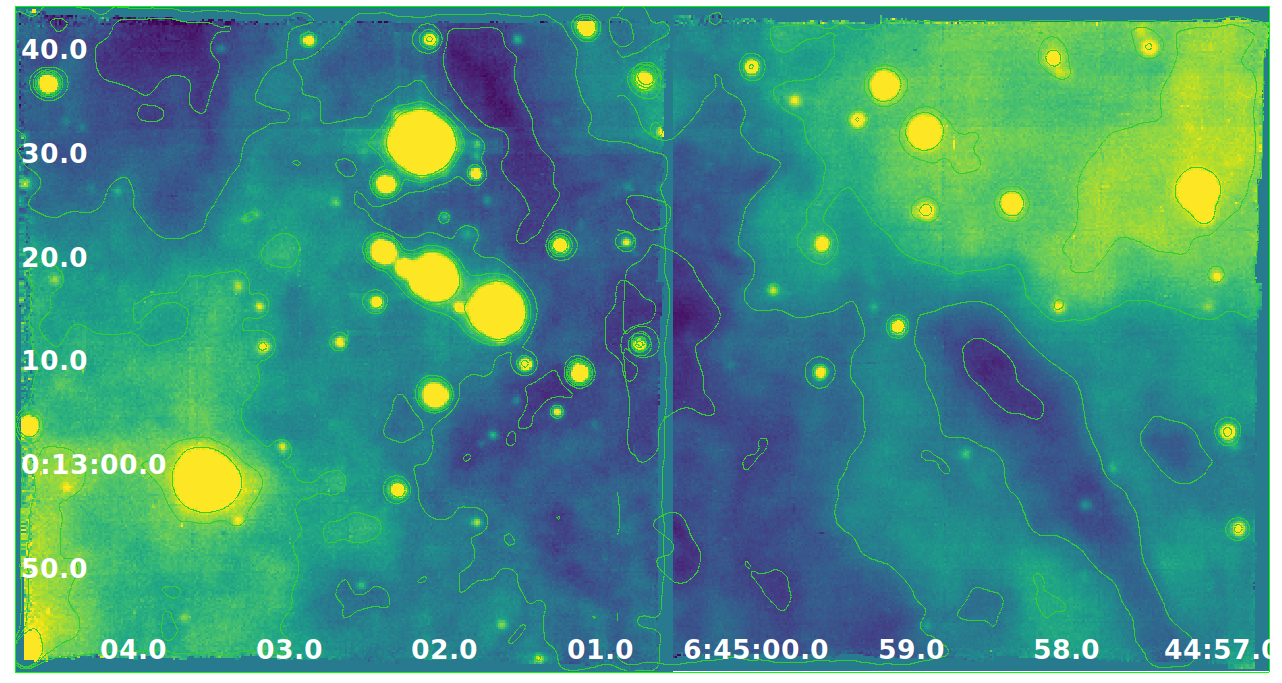}
    \caption{H$\alpha$ intensity map created from the MUSE data cube by integrating over a 10\,\AA\ wavelength range centered on the H$\alpha$ emission line.}
    \label{fig:ha_map}
\end{figure}

The spectral index-derived \(T_{\text{eff}}\) values were compared to those obtained from the continuum fitting method, which employs \(\chi^2\) minimization between observed spectra and the same BT-Settl grid. The two methods show excellent agreement, with a median offset of 25 K and a dispersion of ±200 K (Figure~\ref{fig:si histogram}). 

While the indices provide precise estimates for cool stars \(T_{\text{eff}} \leq 4500\,\text{K}\), their diagnostic power diminishes sharply at higher temperatures, introducing systematic uncertainties for stars near or above this limit. In contrast, the continuum fitting method remains robust across the full \(T_{\text{eff}}\) range of the sample, as it relies on the overall continuum shape, which retains temperature sensitivity even in the absence of TiO. In addition, the continuum-based fitting includes \(A_V\) as a free parameter, providing information on extinction, whereas the silicon lines constrain only the effective temperature \(T_{\text{eff}}\).

\begin{figure}
    \centering
    \includegraphics[width=0.48\textwidth]{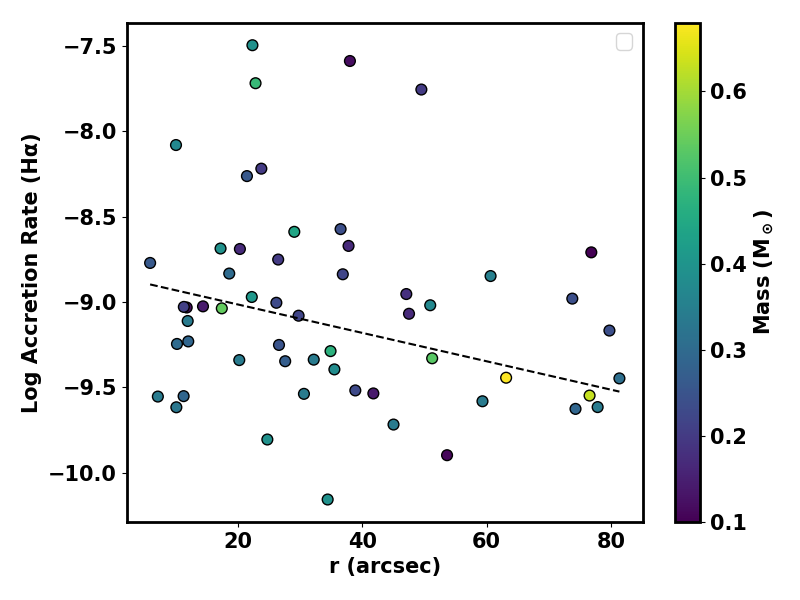}
    \caption{Logarithmic accretion rate derived from H$\alpha$ emission as a function of projected distance ($r$) from the central massive star. Each point represents a young stellar object, color-coded by stellar mass. The dashed line indicates the best-fit linear trend, which suggests a mild decrease in accretion rate with distance. However, given the small field of view and the likely influence of other massive stars outside the observed region, this trend cannot be considered robust.}
    \label{fig:r_vs_ha}
\end{figure}

\begin{table*}
    \caption{Table of parameters for accreting sources with H$\alpha$ equivalent width greater than 10\,\AA.}
    \begin{tabular}{@{}lcccccc@{}}
    RA &   Dec & Mass ($M_\odot$) & Radius ($R_\odot$) & Log Age (Myr) & $\log\dot{M}_{\text{acc}}$ (H$\alpha$) & $\log\dot{M}_{\text{acc}}$ (H$\beta$) \\
    \hline
    101.249 & 0.226 &      0.24 ± 0.03 &        1.85 ± 0.19 &   5.16 ± 0.05 &                            -8.6 ± 0.13 &                          -9.11 ± 0.15 \\
    101.249 & 0.226 &      0.47 ± 0.14 &        0.72 ± 0.07 &   6.94 ± 0.03 &                            -9.3 ± 0.14 &                          -9.31 ± 0.13 \\
    101.248 & 0.226 &      0.23 ± 0.09 &        0.74 ± 0.07 &   6.55 ± 0.04 &                           -9.51 ± 0.30 &                           -9.3 ± 0.21 \\
    101.253 & 0.227 &      0.48 ± 0.08 &        2.51 ± 0.25 &   5.54 ± 0.05 &                           -7.72 ± 0.10 &                          -7.96 ± 0.09 \\
    101.253 & 0.227 &      0.39 ± 0.05 &         0.97 ± 0.1 &   6.46 ± 0.08 &                            -9.8 ± 0.15 &                          -9.36 ± 0.11 \\ 
    \end{tabular}
    \label{table:accretors}
    \tablecomments{Table 1 is published in its entirety in the machine-readable format.
      A portion is shown here for guidance regarding its form and content.}
\end{table*}

\begin{table*}
    \caption{Physical properties of confirmed cluster members.}
    
    \begin{tabular}{ccccccc}
    $\alpha$ (J2000) & $\delta$ (J2000) & $\mathrm{T_{eff}}$ & $A_V$ & $L_{\odot}$ & Mass ($M_{\odot}$) & log(Age) \\
    \hline
    101.263989 & 0.221313 &   3,610.00 ± 32.0 & 0.6 ± 0.27 &           0.21 ± 0.02 &  0.32 ± 0.04 & 6.14 ± 0.08 \\
    101.261884 & 0.225456 &   3,600.00 ± 59.0 & 2.0 ± 0.43 &           0.22 ± 0.03 &   0.3 ± 0.06 &  6.1 ± 0.14 \\
    101.260045 & 0.215339 &   3,590.00 ± 73.0 & 1.8 ± 0.49 &           0.13 ± 0.02 &  0.38 ± 0.05 & 6.49 ± 0.16 \\
    101.268908 & 0.221480 &   3,520.00 ± 55.0 & 2.0 ± 0.39 &           0.47 ± 0.06 &  0.17 ± 0.02 & 4.95 ± 0.19 \\
    101.237431 & 0.218657 &   3,500.00 ± 78.0 & 2.2 ± 0.56 &           0.13 ± 0.02 &  0.31 ± 0.07 &  6.33 ± 0.2 
    \end{tabular}
    
    \tablecomments{Table 2 is published in its entirety in the machine-readable format. A portion is shown here for guidance regarding its form and content.}
    \label{table: all_members}
\end{table*}

\clearpage

\figsetstart
\figsetnum{21}
\figsettitle{Best-fit models for identified cluster members}

\figsetgrpstart
  \figsetgrpnum{21.1}
  \figsetgrptitle{Best-fit models (page 1)}
  \figsetplot{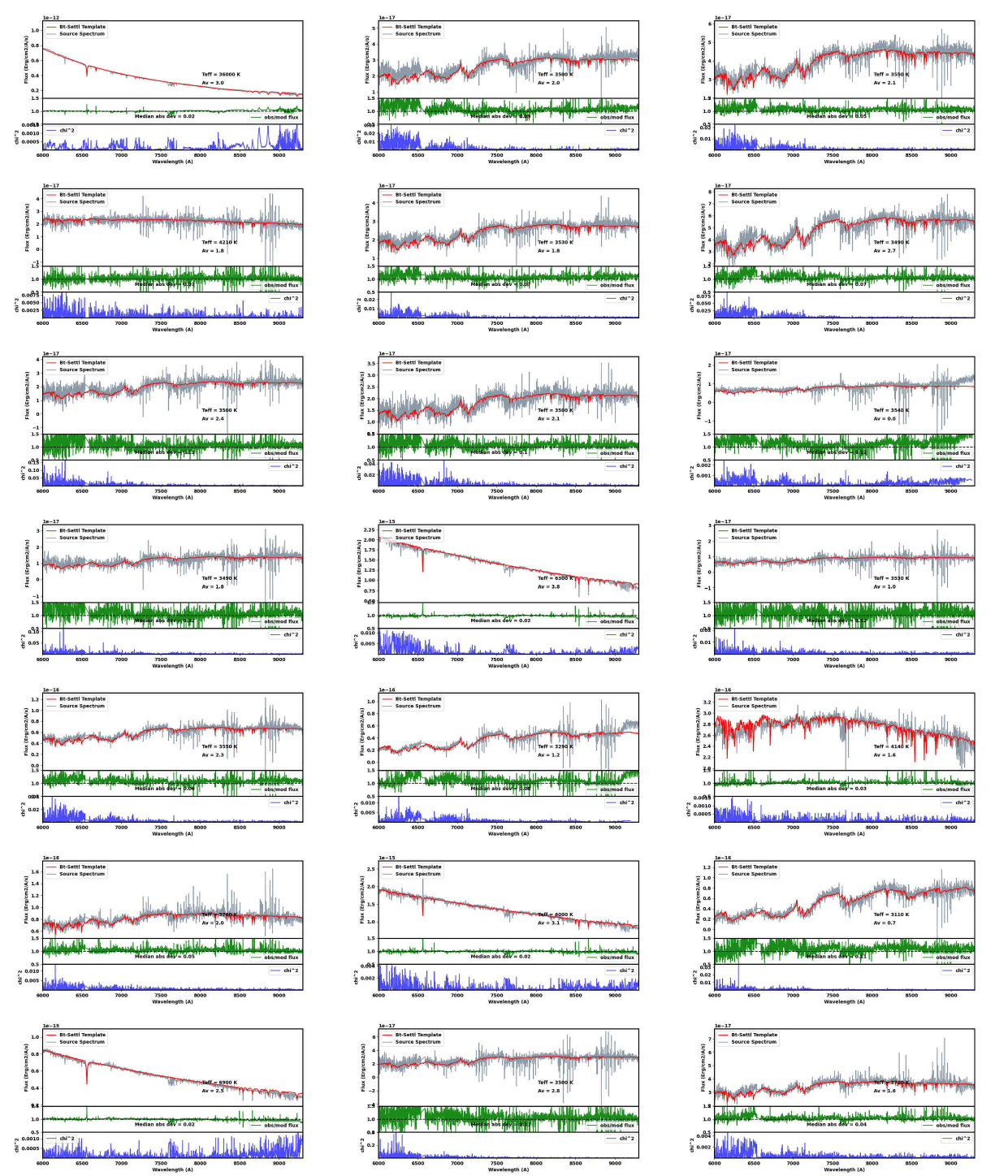}
  \figsetgrpnote{Best-fit spectral models for cluster members (page 1 of 6).}
\figsetgrpend

\figsetgrpstart
  \figsetgrpnum{21.2}
  \figsetgrptitle{Best-fit models (page 2)}
  \figsetplot{bestfit_plots_grid_2.pdf}
  \figsetgrpnote{Best-fit spectral models for cluster members (page 2 of 6).}
\figsetgrpend

\figsetgrpstart
  \figsetgrpnum{21.3}
  \figsetgrptitle{Best-fit models (page 3)}
  \figsetplot{bestfit_plots_grid_3.pdf}
  \figsetgrpnote{Best-fit spectral models for cluster members (page 3 of 6).}
\figsetgrpend

\figsetgrpstart
  \figsetgrpnum{21.4}
  \figsetgrptitle{Best-fit models (page 4)}
  \figsetplot{bestfit_plots_grid_4.pdf}
  \figsetgrpnote{Best-fit spectral models for cluster members (page 4 of 6).}
\figsetgrpend

\figsetgrpstart
  \figsetgrpnum{21.5}
  \figsetgrptitle{Best-fit models (page 5)}
  \figsetplot{bestfit_plots_grid_5.pdf}
  \figsetgrpnote{Best-fit spectral models for cluster members (page 5 of 6).}
\figsetgrpend

\figsetgrpstart
  \figsetgrpnum{21.6}
  \figsetgrptitle{Best-fit models (page 6)}
  \figsetplot{bestfit_plots_grid_6.pdf}
  \figsetgrpnote{Best-fit spectral models for cluster members (page 6 of 6).}
\figsetgrpend

\figsetend

\begin{figure*}
  \centering
  \includegraphics[width=0.99\linewidth]{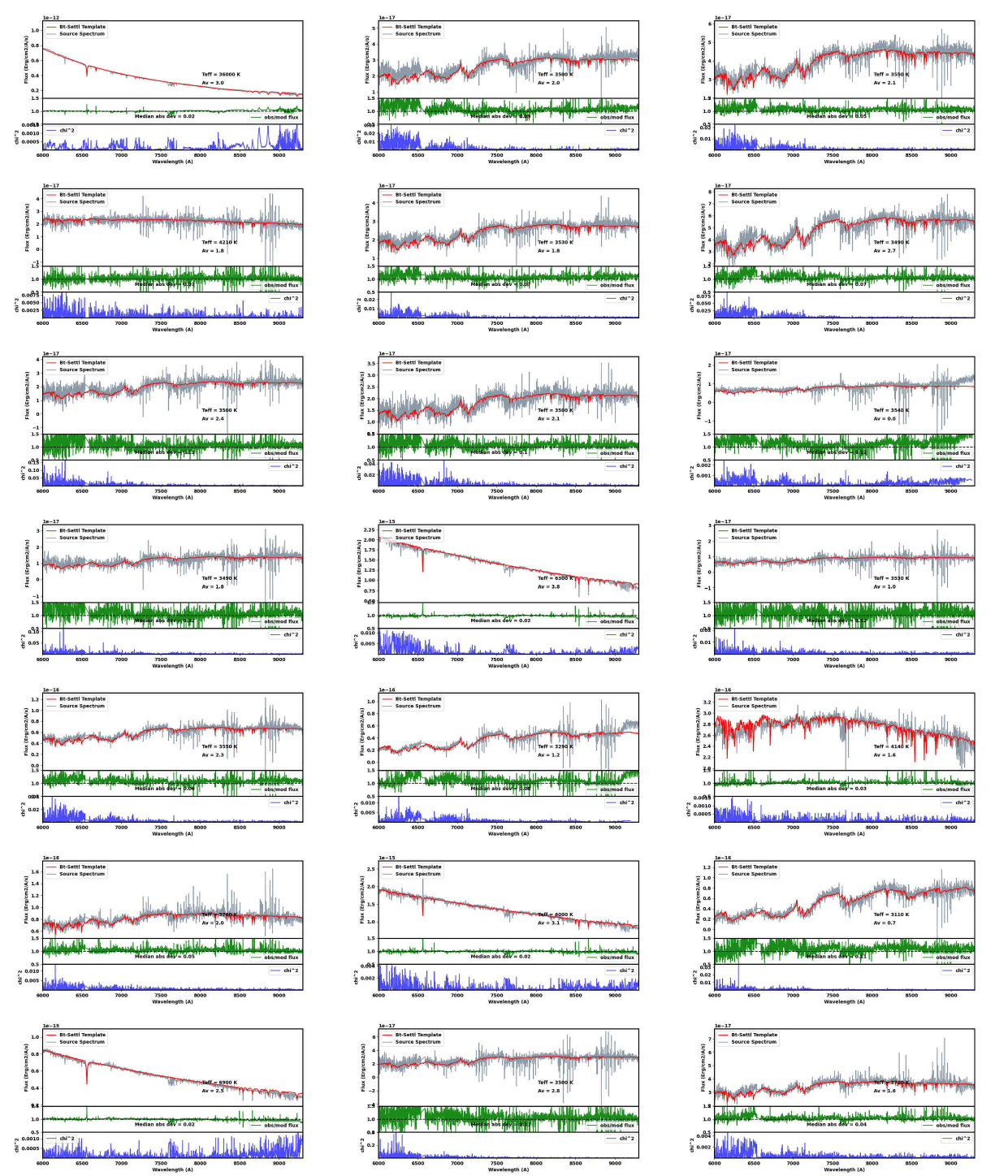}
  \caption{Best-fit models for identified cluster members. 
  The complete figure set is available in the online journal.}
  \label{fig:bestfit}
\end{figure*}

\end{document}